\documentclass[acmsmall]{acmart}

\usepackage{xcolor}
\definecolor{shadecolor}{RGB}{100,100,100}
\usepackage{soulutf8}
\usepackage{wrapfig, booktabs, lipsum, graphicx, soul, xcolor, color}

\AtBeginDocument{%
  \providecommand\BibTeX{{%
    \normalfont B\kern-0.5em{\scshape i\kern-0.25em b}\kern-0.8em\TeX}}}

\setcopyright{acmcopyright}
\acmJournal{PACMHCI}
\acmYear{2021} \acmVolume{5} \acmNumber{CSCW2} \acmArticle{343} \acmMonth{10} \acmPrice{15.00}\acmDOI{10.1145/3476084}

\received{January 2021 }
\received[revised]{April 2021 }
\received[accepted]{May 2021}

\begin{document}

\title[Protection or Punishment?]{Protection or Punishment? Relating the Design Space of Parental Control Apps and Perceptions About Them to Support Parenting for Online Safety}


\author{Ge Wang}
\email{ge.wang@cs.ox.ac.uk}
\affiliation{
  \institution{Department of Computer Science. University of Oxford}
  \city{Oxford}
  \country{UK}
}

\author{Jun Zhao}
\email{jun.zhao@cs.ox.ac.uk}
\affiliation{
  \institution{Department of Computer Science. University of Oxford}
  \city{Oxford}
  \country{UK}
}

\author{Max Van Kleek}
\email{max.van.kleek@cs.ox.ac.uk}
\affiliation{
  \institution{Department of Computer Science. University of Oxford}
  \city{Oxford}
  \country{UK}
}

\author{Nigel Shadbolt}
\email{nigel.shadbolt@cs.ox.ac.uk}
\affiliation{
  \institution{Department of Computer Science. University of Oxford}
  \city{Oxford}
  \country{UK}
}

\renewcommand{\shortauthors}{Ge Wang et al.}

\begin{abstract}
\emph{Parental control apps}, which are mobile apps that allow parents to monitor and restrict their children's activities online, are becoming increasingly adopted by parents as a means of safeguarding their children's online safety.  However, it is not clear whether these apps are always beneficial or effective in what they aim to do; for instance, the overuse of  restriction and surveillance has been found to undermine parent-child relationship and children's sense of autonomy. While previous research has categorised and taken inventory of key features of popular parental control apps, they have not systematically analysed the ways such features were designed or realised in such apps, or in particular how aspects of such designs might relate to parents and children's experiences with such apps.  In this work, we investigate this gap, asking specifically: how might children's and parents' perceptions be related to how parental control features were designed?  To investigate this question, we conducted an analysis of 58 top Android parental control apps designed for the purpose of promoting children's online safety, finding three major axes of variation in how key restriction and monitoring features were realised: granularity, feedback/transparency, and parent-child communications support. To relate these axes to perceived benefits and problems, we then analysed 3264 app reviews to identify references to aspects of the each of the axes above, to understand children's and parents' views of how such dimensions related to their experiences with these apps. Our findings led towards  1) an understanding of how parental control apps realise their functionalities differently along three axes of variation, 2) an analysis of exactly the ways that such variation influences children's and parents' perceptions, respectively of the usefulness or effectiveness of these apps, and finally 3) an identification of design recommendations and opportunities for future apps by contextualising our findings within existing digital parenting theories.
\end{abstract}

\ccsdesc[500]{Human-centered computing~Walkthrough evaluations}
\ccsdesc[500]{Human-centered computing~HCI theory, concepts and models}
\ccsdesc[500]{Human-centered computing~Empirical studies in HCI}
\keywords{Children online safety; parental controls; mobile apps; parental mediation}

\maketitle

\section{Introduction}

Children are spending an unprecedented amount of time online each day via their smartphones and tablets. In the UK, for instance, 96\% of children aged 5-15 are online, and more than half of ten-year-olds have their own smartphones or tablets \cite{ofcom2020}. While the Internet has become an essential enabler for children to learn, have fun, and grow–--especially during a pandemic---there are, of course, significant risks associated with children going online, especially as they engage in new digital activities \cite{eukids2020, policy20218}. Increasing concerns are discussed regarding excessive screen time \cite{st2020, st2019, wang2019concerns}, as well as more direct risks including cyber bulling, inappropriate or harmful content, among others~\cite{zhao2019make, harm2017, pinter2017adolescent}. In response to children's increasing exposure to both classes of risks, a new genre of apps, known as \emph{parental control apps} have emerged. These tools are designed to act as technical mediation support for parents to facilitate  access to, and control over, their children's online activities as a means of protecting them from such harms~\cite{zaman2016parental}. These apps have rapidly grown in popularity over the past few years. The global parental control software market is anticipated to grow from \$1.52Bn USD in 2017 to \$2.53Bn USD by the end of 2023~\cite{mrfr}.

This rapid adoption and increasing reliance on parental control apps has raised corresponding questions about their efficacy --- how such apps fit into existing parenting strategies, and the effects such apps are having on parents and their children.  Only a few empirical studies of these apps have thus far been conducted, and have started to reveal some shortcomings of these apps. Wisniewski et al. (2017) ~\cite{wisniewski2017parental} derived a Teen Online Safety Strategies Framework (TOSS) and applied it for an analysis of features of 75 popular parental control apps for teenagers. A following-up analysis of reviews on these apps by children found many reasons they disliked parental control apps, including the ways they overly restricted them and disenfranchised them of autonomy and privacy, but also identified positive aspects, such as for managing screen time and keeping them from harmful content~\cite{ghosh2018safety}. These studies reveal there may be significant room to improve parental control apps not just towards supporting more effective parental mediation, but ensuring that such measures do not inadvertently introduce new harms or take away valuable opportunities and benefits of digital environments for children. 

To understand where and how such improvements might be achieved, we used Wisniewski et al. (2017) ~\cite{wisniewski2017parental} as the baseline for our research. In this paper we first analysed 58 popular Android parenting control apps to examine how features on the current app market maps to the TOSS framework, and then we further analysed the ways these features were specifically realised. We found three primary axes of feature variation relating to \emph{granularity}, which refers to the level of control an app enables parents to do or the level of information given to the parents, \textit{feedback/transparency}, which refers to the different designs that support varied level of information given to the children, and finally \textit{parent-child communications support}, which reflect how apps supported or stimulated discussions between parents and children about their online activities. These axes of variation offer an orthogonal way of organising the current parental app design space in comparison to TOSS, allowing us to focus on the different designs used to support each function. To then understand how each of these axes relate to how apps effectively support or create problems for children, we conducted a thematic analysis of both children's and parents' reviews regarding how their perceptions varied across the ways features were implemented.

The contributions of this paper are threefold, as follow: 1) first, it extends Wisniewski and Ghosh's ~\cite{wisniewski2017parental} TOSS framework by introducing three orthogonal axes of variation, which contributes an understanding of how parental control apps realise their functionalities differently along these three axes of variation, 2) second, by relating such axes to positive and negative reviews of apps, we identify how the design of features may influence children's and parents' perceptions of them, and finally 3) we identify gaps in the current parental control app design space, contextualising our findings to digital parenting theories. From this, we derive at least two approaches that more autonomy-supportive parental control apps might be achieved.  The first, by simply designing app features to fall along each of the three axes within our identified ``zone of best practice''--designs  perceived most positively by both parents and children.  Second, to complement and pair restriction and monitoring features in an age and skill-appropriate manner, e.g. to gradually allow children to move from strictest monitoring and restriction features to gain more autonomy as they develop necessary skills associated with recognising and mitigating online risks, and to prioritise the use of verbal agreements, thus reframing the use of restriction and monitoring features as a means of scaffolding children's online skills.

\section{Related Work}

\subsection{Children's Online Safety}
Children's online safety has been a broad topic that is closely related to our research. Livingstone et al. ~\cite{livingstone2015framework} categorised the different types of online risks under \textit{content (child as receiver), contact (child as participant) and conduct (child as actor)}, in which children may experience online risks of four types - aggressive (violent content, harassment, and hostile peer activity), sexual (pornographic content, sexual abuse online, `sexting'), values (hateful content, ideological persuasion, potentially harmful user-generated content) and commercial (embedded marketing, personal data misuse, copyright infringement). Similarly, the UK Online Harms White Paper ~\cite{hm2019online} identified the scope of online harms as explicit harms (e.g. harassment and cyberstalking), harms with a less clear definition (e.g. cyberbullying and trolling), and underage exposure (e.g. inappropriate material). A survey done by Ofcom in 2019 showed that data-related harms (e.g. data collection in obscure ways), security harms (e.g. scams/fraud) and content/contact harms (e.g. offensive language) were the most commonly reported online harms experienced by children~\cite{ofcom2019}.

Children's online safety has become an even more critical issue when it comes to their mobile online safety, due to the ubiquitous nature of mobile devices.  Existing online safety problems including stranger danger, online inappropriate content, bullying problems have been amplified. Children now have instant access to the internet~\cite{smahel2020eu,mollborn2020making}, and parents reported that it has been even harder for them to keep track of their children's online activities when their children were using devices of their own~\cite{PEW2020}. Children have been subjected to mobile use starting from a very young age. In their initial years (3-5), children's mobile use are still mainly adult-guided. Several studies indicate that it is not uncommon for parents to rely on technology to keep their children busy or entertained, regularly using a smartphone to replace babysitting  \cite{livingstone2011risks}. Meanwhile, children at this stage do not always understand the difference between fantasy and reality~\cite{pine2002dear, ali2009young}. For slightly older children (6-11), they are gaining increasing independent use of devices \cite{ofcom2017, ofcom2016}. Children at this stage are learning about the complexities of relationships and becoming socially more sophisticated. Contacting their friends through mobile devices become an essential part of their lives, due to the need to fit in and be accepted by the peer group~\cite{kidron2017digital}.  Meanwhile, they are particularly vulnerable to `tech tantrums' - reward loops and auto-plays make it difficult for them to manage their online usage because their evolutionary biology (need to react) is exploited by random rewards and interventions \cite{disruptedchildhood}. For children between 12-15, phone becomes a key social information and education tool. They start to become more heavily involved in social media platforms through their devices. A recent report showed that by the age of 13 already (the minimum age to open an account on many social media platforms) , more than half of them have a social media profile~\cite{ofcom2019}. Meanwhile, children at this stage are undergoing significant neuro-psychological changes, leading to differences in how they perceive emotions and make decisions~\cite{kidron2017digital, paulsen2012risk}. They often become easily influenced by the content they see as they tend to be characterised by idealism, with a tendency towards polarised thinking \cite{temple1997cognitive, briones2013kony}. Today's children are exposed to the same range of online risks on mobile platforms, only with risks becoming even more exacerbated and pervasive.

\subsection{Theoretical Frameworks For Children's Online Safety}
Safeguarding children during their use of technologies has been a long concern for parents. With more of these theories being embedded in technical designs and empirically evaluated, we are seeing a transition of focus from the parent-centred approach that emphasises on restrictions and family rules, to the more child-centred approach that encourages self-autonomy and empathy. 
\subsubsection{Parental Mediation Theory: From Pre-Internet Era to Digital Parenting} The basis of work focusing on ways that parents can regulate their children's media habits is grounded in parental mediation theory. Pre-dating the Internet era, the focus of parental mediation was on children's TV consumption \cite{nathanson1999identifying, valkenburg1999developing} and how parents can prevent or mitigate harmful effects of children's exposure to increasing amounts of TV content, in particular, that containing violent and inappropriate themes. It set up three mediation strategies: \emph{restrictive mediation}, setting rules and regulations about children's television viewing, \emph{co-view}, simply watching television with children, and \emph{active mediation}, or talking with children about the content they saw on television. With increasing concern over risks beyond media exposure, parental mediation theory has been adapted to incorporate these broader concerns regarding children's online safety. It was refocused to the goal of minimising online risks in general while maximising opportunities for children \cite{livingstone2008parental, mesch2009parental}. This change of context required considerable refinements to the constructs of existing theories. Restrictive mediation, for instance, was recast to refer to parents setting the family's rules and boundaries around online activities. While the restriction of television media consumption was relatively easy as parents can take guidance from television program classification \cite{clark2011parental}, online restriction requires much more effort from parents as it involves a great more of diversity. This modern version of restrictive mediation now also includes some technical means, such as using software restrictions to limit access to content or screen time \cite{nikken2014developing}. Active mediation is still centered around parent-child communication and refers more specifically to parents' active efforts to guide, interpret, and discuss online harms and risk copings with children \cite{clark2011parental}. Co-view has been replaced by a more ``internet-specific'' strategy - monitoring children's activities online.

\subsubsection{Teen Online Safety Strategies Framework (TOSS)}
\begin{figure}[h]
    \centering
    \includegraphics[width=1\textwidth]{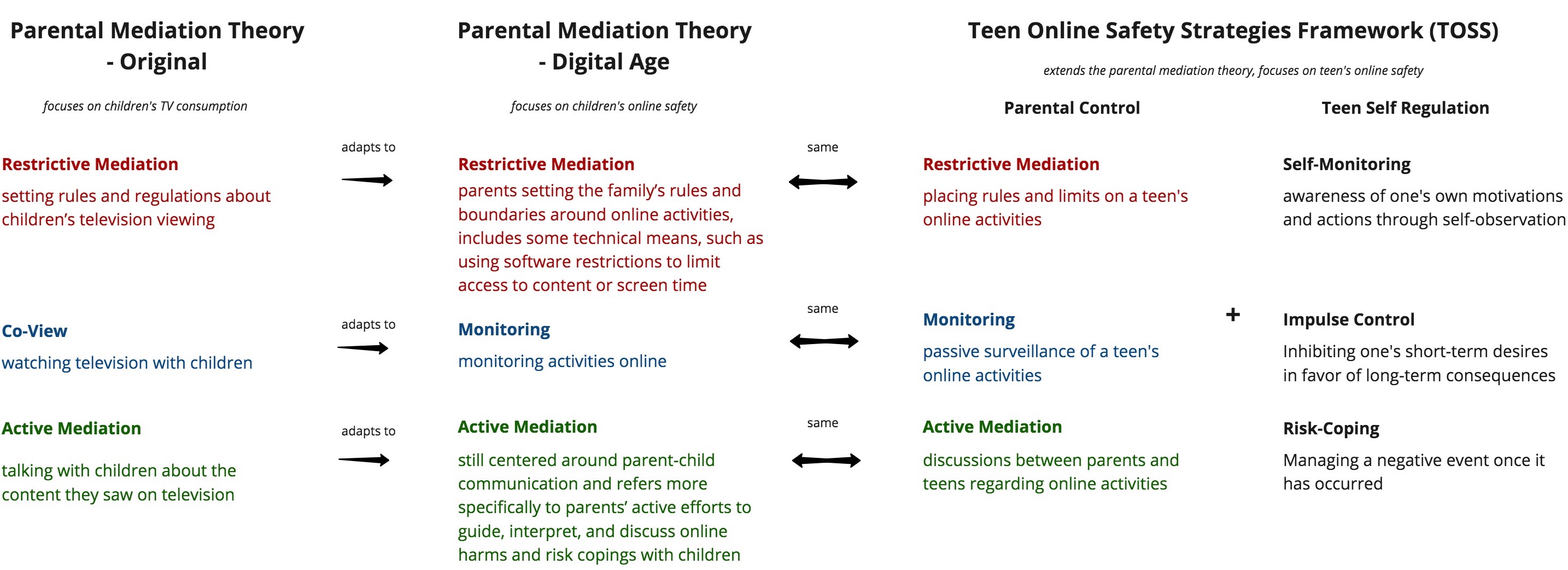}
    \caption{How parental mediation theory changes through the digital age and how it connects to the TOSS framework}
    \label{fig:pmt}
\end{figure}

The TOSS (Teen Online Safety Strategies Framework) proposed by Wisniewski et al. \cite{wisniewski2017parental} made a novel attempt of highlighting the gap between a parent-centred and a child-centred mediation approach, see Figure~\ref{fig:pmt}. TOSS is a theoretically derived conceptual framework of teen online safety strategies, comprised of two main strategies - parental control, which was originally derived from parental mediation theory; and teen self-regulation -  the framework extends the mediation theory by introducing a complementary set of strategies centred around the child/teenager orientated around teen self-regulation.  

As shown in Figure~\ref{fig:pmt}, the parental control strategies contain three sub-strategies, including \textit{monitoring}, \textit{restriction}, and \textit{active mediation}. The teen self-regulation strategies comprised \textit{self-monitoring}, in which children self-regulate their behaviors through self-observation; \textit{impulse control}, which refers to children exhibiting their short-term desires in favor of the long-term consequences; and \textit{risk-coping}, defined to be a self-regulatory process that occurs after one found themselves in risks, which typically involves processes such as problem-solving, advice-seeking and acquiring social support. 

Although originally developed for analysing \textit{teen} online safety strategies, the TOSS framework is equally valuable for analysing \textit{children} online safety strategies, especially the \textit{parental control} side of the framework. Furthermore, the framework provides a useful way for analysing parental control apps practices. In the original paper the framework was proposed, the authors used the TOSS framework to analyse features that support parental control strategies and teen self-regulation strategies \cite{wisniewski2017parental}, and they found that the apps strongly favored features that promote parental control through monitoring and restricting teens online. The framework provides the conceptual basis of our work here, which relies on both Parental Mediation Theory and TOSS. In particular, we used the TOSS framework as the baseline for our app analysis.

\subsubsection{Mediation Strategies: A Dynamic Process}
\label{subsubsection:beyondTOSS}
Both the parental mediation theory and TOSS categorised the online safety strategies into distinct types. Previous research argued that the parental mediation theory tends to address restriction, monitoring, and active mediation strategies independently of each other \cite{beyens2017parent, meeus2018managing}. However, several studies have challenged the view that one mediation type should be significantly better or worse than the others. They instead found that parent-child conflicts depend more on the styles in which parents provide mediation, regardless it's restriction, monitoring, or active \cite{beyens2017parent, valkenburg2013developing}. For example, Valkenburg et al~\cite{valkenburg2013developing} suggested that active, monitoring or restrictive mediation strategies are not in themselves good or bad; their effectiveness may depend on the style in which parents apply these strategies to their children. Examples of such styles were adapted from a self-determination theory viewpoint \cite{ryan2000self}, namely \textit{controlling}, parents pressuring children to think and behave in certain ways through, for example, guilt induction or criticism \cite{fikkers2017matter}; or \textit{autonomy-supportive}, which refers to ``structure and guidance but takes the child’s feelings and perspective seriously providing a convincing rationale for behavioural requests and rule-making'' \cite{valkenburg2013developing}; or \textit{inconsistent}, in which parents were sometimes strict and at other times acquiesce to their children \cite{fikkers2017matter}.

Similarly, studies on families showed that parents do not use active and restrictive mediation strategies independently of each other. Instead, parents who were motivated to play an active part in their children’s online activities also tend to implement monitoring and restriction through various active ways \cite{chen2016active}. These findings echo similar work that suggested that parental mediation may be better conceived as a dynamic process that should integrate into the daily interactions between parents and children, rather than as a set of distinct strategies implemented separately \cite{symons2017parental}. 

\subsubsection{Online Safety as Skill Development}
Parental mediation theory can be dated back to developmental psychology research in the early 20th century on how children's developmental could be guided by interpersonal communication ~\cite{reynolds1971mutually}. While parental mediation theory drew upon Bandura's social learning theory ~\cite{bandura1977social} that posit parents as `role models' and explore how parental involvement may mitigate the negative behaviour of children. There has been another line of work that also originated from interpersonal communication studies, but place at the center of the investigation the children and their interactive experiences with adults. Vygotsky ~\cite{chaiklin2003zone} theorised children's potential for cognitive development using the term `zone of proximate development'. This `zone' describe the things that's beyond the children's current capabilities, but is `just about' to be grasped by them as long as they are given sufficient `scaffolding' - guidance and support from more knowledgeable others. A survey of 215 parents and their teens showed that increased parental control was associated with more (not fewer) online risks~\cite{10.1145/3173574.3173768}. On the other hand, a study of 18 U.S. families with children ages 5-11 found that children largely relied on their parents for online risks support, and parental scaffolding via conversations and interactions was an essential part of their rick coping skills development~\cite{kumar2017no}. Similarly, a study with children aged 6-10 in the UK showed children were well capable of grasping the key ideas of online risks and develop coping skills, when sufficient scaffolding is given~\cite{zhao2019make}.

\subsection{Parental Control Apps For Children's Online Safety}
With the increase of digital technologies in families' life, parental controls mobile apps are commonly used by parents to mediate their children's online usage. Research has been trying to assess how effective these apps are for safeguarding children online. A review of 75 parental control apps \cite{wisniewski2017parental} showed that restriction and monitoring approaches were most common across the commercial market of these parental control apps. A further study on users' perspectives found that children mainly felt these apps were overly restrictive and invasive of their privacy and negatively impacting their relationship with their parents \cite{ghosh2018safety}. Recent research also explicitly analysed the linguistic patterns of parent's reviews versus children's reviews on parental control apps and found that tensions were more likely to occur when apps failed to solve problems regarding children's safety \cite{alelyani2019examining}. Apart from users' perceptions, existing literature also suggested there are significant privacy concerns around these apps. In a study done on 46 parental control apps, researchers found that 72\% of them share data with third parties without mentioning their presence in privacy policies \cite{feal2020angel}. 

Following these understandings, researchers are beginning to explore new ways to enable this safeguarding. For instance, Ghosh et al.~\cite{ghosh2020circle} proposed \textit{Circle of Trust}, a parental mediation tool that integrates positive family values  and aims to strike a balance between teen's privacy and their online safety. Hashish et al.~\cite{hashish2014involving} designed We-Choose, a tool for controlling content on the smart tablets for children between six to eight years. It supports collaborative rule-setting by facilitating discussion on the appropriateness of apps and helping children to review their choices . Ko et al.~\cite{ko2015familync} developed \textit{FamiLync}, a tool that facilitates participatory parental mediation by encouraging both parents and children to participate in co-learning of digital media use. However, we still lack a systematic understanding of how various parental mediation theories are currently realised or designed in the commercial market.

\section{Methods}

\begin{figure}[h]
    \centering
    \includegraphics[width=1\textwidth]{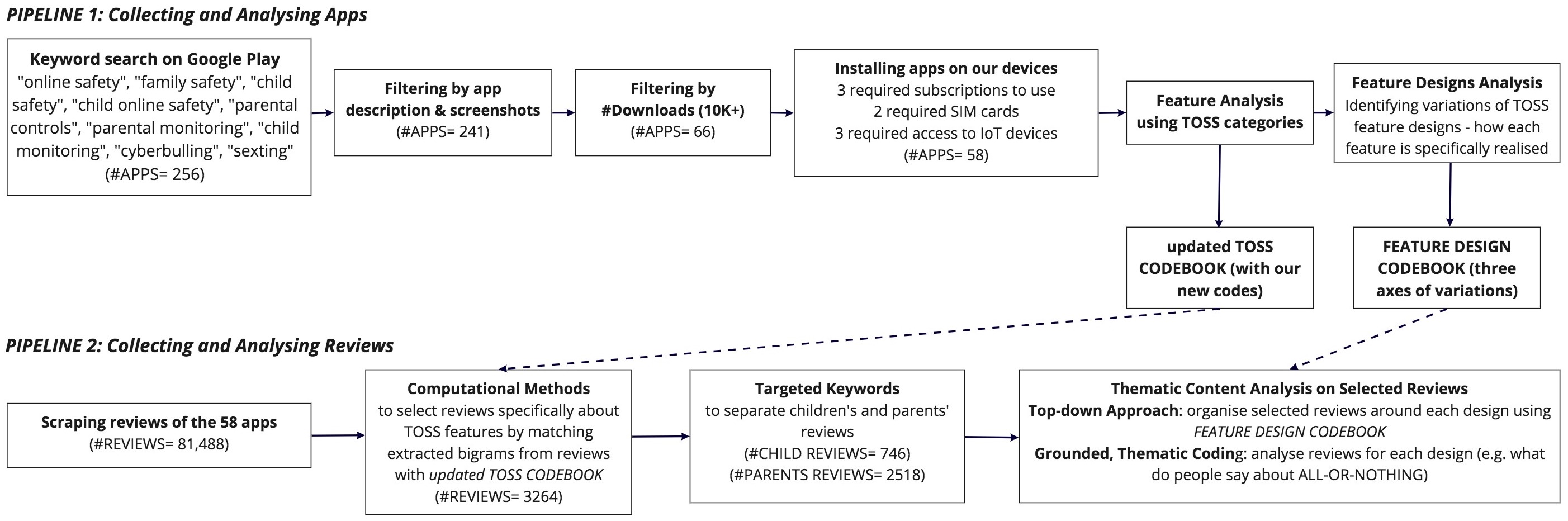}
    \caption{Our methodology: we started by replicating the method by Wisniewski et al~\cite{wisniewski2017parental} and used their results as baseline for our analysis. We then extended our data analysis beyond their methodologies by 1) analysing how the app features are realised; and 2) conducting a thematic analysis on parents' and children's reviews of these features in order to identify their positive v.s. negative perceptions. }
    \label{fig:pipeline}
\end{figure}

We used two pipelines to explore our research questions (see Figure~\ref{fig:pipeline}), to analyse app feature designs and their app reviews. Through this, we wanted to review how parental control is currently realised to identify gaps in the market and further see how these features are perceived by users. 

We particularly aligned our app feature coding with the qualitative feature analysis conducted by Wisniewski et al. in 2017~\cite{wisniewski2017parental}. Their analysis included 75 Android mobile parental control apps and led to an identification of 42 app features (e.g. website monitoring, or app activity blocking) that are mapped to the TOSS conceptual framework (see Figure~\ref{fig:pmt}). Our methodology was based on and extended the work by Wisniewski et al~\cite{wisniewski2017parental}. We started by replicating their method for app data acquisition and then introduced particular analysis about how each of the feature was realised in the current apps. Our intention to replicate here was not just to establish a baseline for our further analysis but also because replication is especially valuable in HCI to analyse quickly evolving phenomena, especially the design of mobile apps. 

\subsection{Collecting Apps}

In July through August of 2020, we sought to identify popular Android parental control applications (“apps”) that promote children's online safety on Google Play. In order to effectively compare our analysis with Wisniewski et al.'s~\cite{wisniewski2017parental} analysis of app features in 2017, we adopted their search keywords (since our research focused on children instead of teens, we replaced the words ``teen'' and ``adolescent'' with ``child''). We performed keyword searches using the terms  ``online safety,” ``family safety,'' ``child safety,'' ``child online safety,” ``parental controls,'' ``parental monitoring,'' ``child
monitoring,'' ``cyberbullying,'' and ``sexting.'' Similar to the original method used by Wisniewski et al~\cite{wisniewski2017parental}, we also included all ‘similar apps’ that were suggested by the search results, and read  each app description \& app screenshots to ensure returned results met the inclusion criteria, which was that the app had to be designed to support the parental mediation of children's online activities.

We generated an initial list of 241 apps. During our process of finding and filtering through these apps, we found that the popularity (in number of downloads) of these apps assumed a long tail distribution: nearly 80\% of these 241 apps had fewer than 20 downloads, while the more popular apps had multi-thousands of downloads.  Since our intentions were to examine features and design elements in the most popular apps, we kept the top 66 apps with at least 10K+ downloads. Among these 66 apps, we further removed the following: three that required subscriptions to use, two that required SIM cards (which was not installed on the test device), and three focused on parental control of IoT devices.  Our final app list included 58 apps. Out of these 58 apps, only 27 of the apps were identical to Wisniewski's original app list in 2017. The full list of apps can be found in Appendix. 


\subsection{Analysing App Features}

In analysing apps, our goals were twofold: to identify key features that these apps have, and, second, to identify whether such apps varied in terms of the ways that such features were realised.

We first generated descriptions of app features by applying a walkthrough method~\cite{light2018walkthrough}, first role-playing as a parent and then as a child using the app. This method is similar to the more commonly used cognitive walkthroughs (CWs) in usability evaluation~\cite{lewis1997cognitive}, in that it is an exploratory screen-by-screen navigation through each app, role-playing as a particular kind of user.  However, there are two important differences: unlike cognitive walkthroughs, our objective was not to identify usability problems, but to identify and descriptively characterise the features of each app. Second, instead of starting with extremely concrete tasks, we started with the goal of approaching each app as a methodical new user might, trying all features provided and systematically exploring their options.

To do this, we installed each app on our devices, in turn.  For parenting apps that came in pairs (corresponding to a parent child version), we installed the parent app on an Android tablet, and the corresponding child app on an Android mobile phone. For those apps that were singular apps, we set up the app in parent mode on the tablet, and installed set up the app in child mode on the mobile phone. The tablet used for analysis was a Huawei MediaPad M5 with 32GB of storage running Android 9 and the phone we used was a Samsung Galaxy S20 with 16 GB of storage running Android 9.

Then, two researchers role-played a parent and then a child, on each respective app, walking through the app a screen at a time to identify features each app provided.  Each feature was examined to identify all options and functionality. Then, a short textual description was generated for each.

\subsubsection{Identifying Distinct Features Using TOSS Categories}
\label{subsubsection:methodTOSSfeature}- We first analysed these descriptions using a ``top-down'' approach, applying the codebook from Wisniewski et al~\cite{wisniewski2017parental}. Although TOSS focused on teen self-regulation, we felt it still appropriate as a starting point for two reasons; first, most such apps did not explicitly designate use by teens only (vs younger children); second, the emerging literature on online safety of younger children suggested that establishing safety-related skills and self-regulation was increasingly pressing in children's younger years due to their increased activity online~\cite{kumar2017no}. Therefore, we chose to interpret the \textit{teen self-regulation} strategies of the TOSS framework as self-regulation strategies for all children \cite{wisniewski2017parental}. Two researchers coded each feature description against the TOSS feature codebook. If a particular feature did not correspond to an existing code, we made a note with a description of its functionality and added it as a new separate feature into our code book. Cohen's kappa was calculated to be high ($\kappa=0.83$) on a hold out sample of 20\% of the apps.

\subsubsection{Feature Design Analysis} \label{subsubsection:methodFeatureDesign} - To explore feature variation, we revisited the feature descriptions and performed a second round of coding, focusing on \emph{how features was realised}.  This analysis phase was, again, performed by two researchers who sub-divided the apps. 
Codes were reconciled, refined, and combined, and then affinity-clustered based on across all TOSS strategies. This process yielded 22 clusters corresponding to three axes: 6 along \emph{granularity}, 8 along \emph{feedback/transparency}, and 8 along \emph{parent-child communications support}, as we describe in detail in Section \ref{subsection:threeaxes}.

\subsection{Collecting App Reviews}
The second analysis pipeline focused on app user reviews; to complement our app analysis, we sought to collect user reviews of each of the 58 apps to understand users' opinions. We scraped all reviews for 58 apps (including the \textit{child app} versions) using open source scraping tools \cite{gps}. This resulted in an initial data set of 93,404 reviews from all 58 apps. Duplicate reviews were then removed, yielding 81,488 reviews.

\subsubsection{Selecting app reviews pertaining to specific features}

Due to the immense number of reviews being spams, irrelevant, or simple statements without justification like ``It's good'', we sought to keep only user reviews that expressed a view about specific TOSS features \footnote{By TOSS feature here, we are not referring to the original features as in Wisniewski 2017, but the features we identified using the TOSS framework (with newly identified features). See Figure \ref{fig:feature_compare} for a full list of our TOSS features.} with some sort of justification. Thus, we developed the NLP pipeline to achieve this.

Our pipeline was based on that by Guzman et al. \cite{guzman2014users}. An initial data processing procedure was done to only keep the nouns, adjectives and verbs in the reviews.  We then used the bigram finding algorithm provided by the NLTK toolkit \cite{bird2009natural} for extracting user reviews around specific features (A bigram is a two-word phrase that co-occurs unusually often). We filtered the bigrams by only considering those that appeared more than five times and had less than three words distance between them. We then clustered bigrams whose pairs of words were synonyms using Wordnet \cite{miller1995wordnet}.  After that, we manually reviewed each bigrams cluster to map it to each TOSS feature (identified in Section~\ref{subsubsection:methodTOSSfeature}). For example, both bigrams clusters  <limit time> and <limit screen> would belong to the SCREENTIME-BLOCK feature, which is about limiting screen time. This gave us an indication whether an app feature was indeed discussed in a review, and hence made it a potentially `meaningful' review. 

We then traced back to the original reviews where these bigrams that can be mapped to a TOSS app feature were extracted from. The author then went through all of these reviews to manually verify that the corresponding TOSS feature was indeed mentioned in the review, remove any inconsistent ones and the ones not about a specific feature. We ended up with a review data set consisting of 3,264 reviews in total. 

To identify perception difference between children and parents, we used a similar method to Ghosh et al~\cite{ghosh2018safety}. This means that we also used certain keywords and phrases to identify posts by parents and children respectively. Examples indicating a child's perspective included: ``my parent'', ``my mom'', ``my dad'', ``I’m a kid'', ``I’m xx years old''. The two authors manually reviewed the automatically generated results. Out of the 3,264 reviews, we identified 746 child reviews and 2,518 parent reviews. We now have a data set consisting of review around each TOSS features, from children, and parents, respectively.

\subsection{Analysing App Reviews}
We conducted a thematic content analysis on the final filtered set of reviews to identify what perspectives parents and children hold in terms of the apps and their features. So far, the review dataset has been labelled with the TOSS app feature (e.g. monitoring website) discussed by each review (for filtering meaningful reviews), and the type of user contributing the review (parent vs. child). Here, we would like to first identify the feature designs discussed in each review, and then analyse the perceptions from parents and children about each feature design --- positive or negative. 

First, the researchers used a top-down approach to read through each review to categorise reviews around feature designs (i.e. how a feature is implemented) we previously compiled in Section \ref{subsubsection:methodFeatureDesign}. Then, for reviews around each type of feature designs (e.g. ALL-OR-NOTHING), we used a grounded, thematic approach~\cite{wildemuth2016applications} to identify themes about why a parent or a child liked or disliked that feature design. (The final codebook for reviews data can be found in Appendix)

The thematic coding process was performed by two researchers who sub-divided the reviews. The hold out sample was 50 reviews (roughly around 35\%) for each feature design (e.g. coarse granularity, low feedback/transparency), with a Cohen's kappa of 0.77.

\section{Results}

\subsection{App Feature Analysis}
\label{subsection:TOSSfeature}

\begin{figure}[h]
    \centering
    \includegraphics[width=1\textwidth]{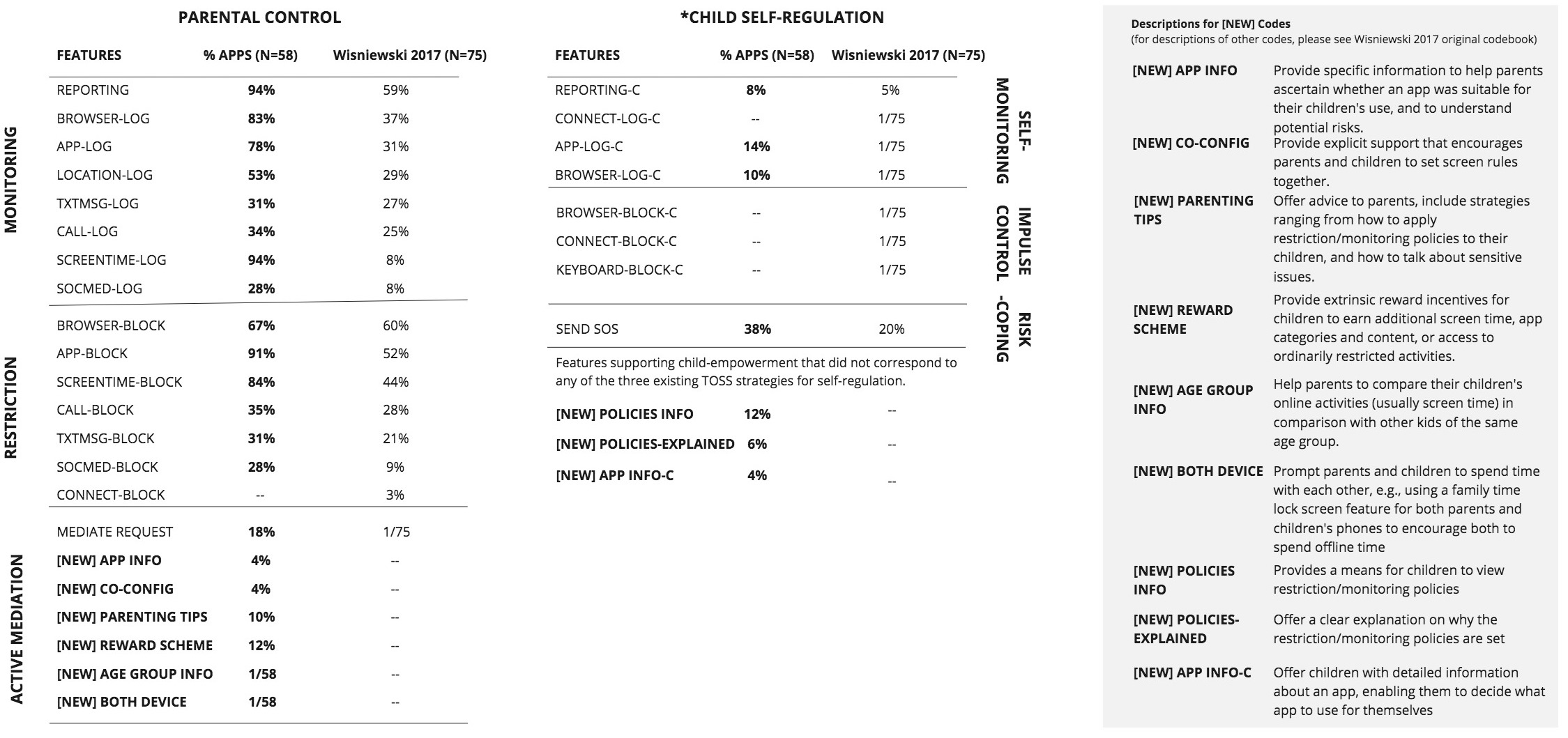}
    \caption{Comparing our TOSS feature analysis results with Wisniewski 2017. We found that in practice, active mediation features were often grounded in monitoring and restriction features; Also, restriction and monitoring features came in a great variety of details that was not captured by TOSS (*Previously \textit{TEEN SELF-REGULATION} in Wisniewski 2017 original code book)}
    \label{fig:feature_compare}
\end{figure}

Applying the TOSS framework~\cite{wisniewski2017parental} to categorise app features, we identified many features that supported both \emph{Monitoring} and \emph{Restriction} strategies. Figure \ref{fig:feature_compare} juxtaposes the number of occurrences of such features we found in our 58 apps against those reported in the original TOSS paper~\cite{wisniewski2017parental}.  As visible in the figure, the prevalence of restriction and monitoring features in apps generally increased compared to the original analysis, suggesting that parental control apps have become more featureful since 2017. 

Moreover, we identified several new features for supporting \emph{Active Mediation} than in the original study. This includes providing information about apps to help parents decide whether an app is suitable for their child (\textbf{APP-INFO}), and features that encourage parents and children to set technology usage rules together (\textbf{CO-CONFIG}). These features are described in the box in Figure \ref{fig:feature_compare}. In terms of teen/child self regulation, we found several features supporting \emph{Self-Monitoring}, including sending reports to children regarding their mobile usage (\textbf{REPORTING-C}), sending app activity logs to children (\textbf{APP-LOG-C}), and sending browser histories to children (\textbf{BROWSER-LOG-C}).In line with prior work, we did not find any features belonging to \emph{Impulse Control}, or any meaningful \emph{Risk-Coping} strategies (apart from sending SOS requests). 

During the analysis process, we found considerable variation among the designs of how the features that belonged to a single code were realised.  For instance, among apps that offered functionality logging children's app usage (\textbf{APP-LOG}), some apps recorded highly granular detail of the child's activities within, and across, particular apps, whilst others only logged events of a specific type, such as `suspicious activities'. This variation was what inspired us to examine the design axes of variation which we describe in section \ref{subsection:threeaxes}. 


Third, we found some features supporting child-empowerment that did not correspond to any of the three existing TOSS strategies for self-regulation. Examples included features designed to provide a means for children to inspect their restriction/monitoring policies (\textbf{POLICIES INFO}), as well as features that offer children with detailed information about an app, enabling them to decide what app to use for themselves (\textbf{APP INFO-C}).

Finally, with respect to app features supporting \emph{Active Mediation}, we found that rather than being standalone, such features were often contextualised within other restriction and monitoring features. Examples included messaging features that let children request more screen time, or access to a restricted app.  Like these, most forms of mediative support concerned restrictions or monitoring activities.  Another type of support included interfaces that encouraged parents to collaboratively discuss and negotiate restriction and monitoring policies with their child (\textbf{CO-CONFIG}), instead of encouraging parents to set such policies without their child's involvement (e.g., \textbf{SCREENTIME-BLOCK}). Even  (\textbf{PARENTING-TIPS}), the closest to being a standalone active mediation feature, was most often associated with tips around restriction or monitoring strategies.

\subsection{Feature Design Variations}
\label{subsection:threeaxes}
\begin{figure}[h]
    \centering
    \includegraphics[width=1\textwidth]{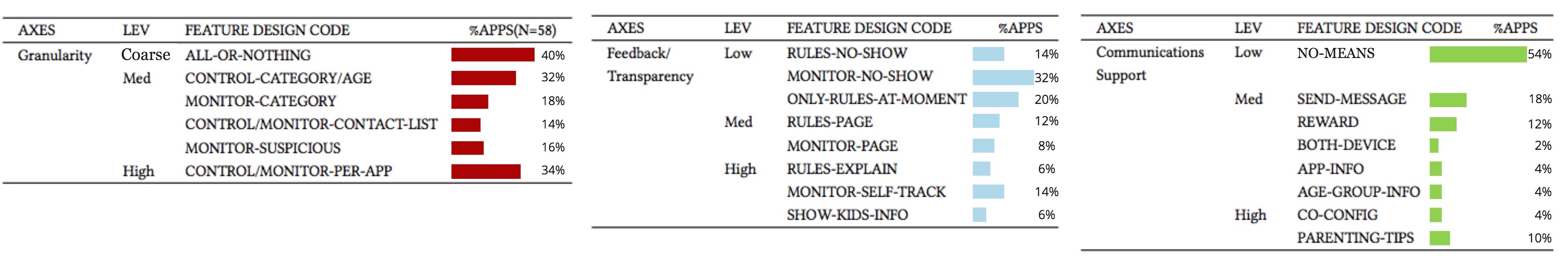}
    \caption{Three axes of variation along which app feature designs varied.}
    \label{fig:designs_stats}
\end{figure}

When looking at how features are specifically realised we identified that they fell into three axes of variation, including \textit{granularity}, which refers to the level of control an app enables parents to do or the level of information given to the parents, \textit{feedback/transparency}, which refers to the different designs that support varied level of information given to the children, and finally \textit{parent-child communication support}, which reflect how apps supported or stimulated discussions between parents and children about their online activities. These axes of variation offer an \textit{orthogonal} way of organising the current parental app design space in comparison to TOSS, allowing us to focus on the different \textit{designs} used to support each function. We identified 6 feature \textit{designs} along granularity, 8 feature \textit{designs} along feedback/transparency, and 8 feature \textit{designs} along communications support that reported that variation of designs of TOSS features, as summarised in Figure~\ref{fig:designs_stats}.

\subsubsection{Axes 1: Granularity}

We observed that the level of control/information provided by apps for the parents spanned from very coarse level of granularity, e.g. control/monitor based on an \textbf{ALL-OR-NOTHING} filtering --- parents having to either block all contents or nothing at all; to highly granular feature designs including allowing parents to configure control/monitor on a per app basis (\textbf{CONTROL/MONITOR-PER-APP}).

\begin{enumerate}
    \item (\textit{Coarse}) ALL-OR-NOTHING : This is a most widely supported design (appearing in 40\% of all apps) that spans across all TOSS features. Parents had to either block all contents and request access to every detail (e.g., every video watched, website visited, the apps opened, even logging google search \&keyword queries.) about children's online activities, or gaining no access about children's activities at all, leaving no middle ground in between. 
    \item (\textit{Medium}) CONTROL-CATEGORY/AGE: In comparison to the all-or-nothing approach, this category allows parents to control children's access based on app/website categories. Such categories were usually derived directly from app store listings or website ratings, and seen to afford parents the convenience of setting broad policies without concerns over specific apps or websites. 
    \item (\textit{Medium}) MONITOR-CATEGORY: Provide only high-level summaries of activities children performed on the phone, such as a list of top contacts, apps used, and time spent on the device per day. These high-level summaries were sometimes grouped by app category.
    \item (\textit{Medium}) CONTROL/MONITOR-CONTACT-LIST: At this level, parents are given a chance to provide customised control by providing a list of a pre-approved contact (or ``suspicious'' contact list); and when children contact people on that list, parents will be acknowledged (or showed with texts etc.) 
    \item (\textit{Medium}) MONITOR-SUSPICIOUS: Similarly, this control of suspicious content can be automatically achieved by apps, which report and alert parents only based on ``suspicious'' or ``dangerous'' activities, messages, or content.
    \item (\textit{High}) CONTROL/MONITOR-PER-APP: At the highest granularity of control, parents are required to configure settings on a per-app (or website) basis.
\end{enumerate}


\subsubsection{Axis 2: Feedback/Transparency}
By feedback/transparency, we refer to the designs (specifically of monitoring features) that support variation of the level of information given to children. While being closely related to the self-monitoring strategies in TOSS framework, this axis extended and enriched it by adding design variation details and discussed designs that supports children's autonomy online. The variation of these designs ranges from very low feedback/transparency - providing no information on the screen rules and things being monitored, to high feedback/transparency - supporting children with resources (e.g. expert reviews, ratings) to let them decide for themselves what to use.

\begin{enumerate}
    \item (\textit{Low}) RULES-NO-SHOW: A considerate amount of apps (14\%) provided no indication about the restrictions enforced on children's phones, leaving them with no idea of the things they could still do on their phones.
    \item (\textit{Low}) MONITOR-NO-SHOW: Nearly a third of the apps in our data set did not acknowledge children about how their information were being monitored by their parents.
    \item (\textit{Low}) ONLY-RULES-AT-MOMENT: This design shows prompts \textit{at the moment} e.g. when children were attempting to access a website on blocking list. However, children weren't informed with the screen rules in advance.
    \item (\textit{Medium}) RULES-PAGE, provides a means for children to view/inspect the restrictions policies, giving them more transparency than the designs above. 
    \item (\textit{Medium}) MONITOR-PAGE, provides some rudimentary information to children about which activities were being monitored, such as browsing history, app use history, device use time, or messaging.
    \item (\textit{High}) RULES-EXPLAINED, identified in a small number of apps reviews (6\%), offers a clear explanation for children when an activity or action exceeded or violated a restriction and why that might be bad for children (rather than simply terminating the activity or giving a generic system error or a blank screen).
    \item (\textit{High}) MONITOR-SELF-TRACK, provides children feedback about their activities as they used the device, such as how much time they had left (total screen time or on an app).
    \item (\textit{High}) SHOW-KIDS-INFO, again identified in only 6\% of all apps, offers children with detailed information of each app, including expert reviews and ratings, enabling children to decide what apps to use for themselves.
\end{enumerate}


\subsubsection{Axis 3: Communications Support}
As this dimension is pertained to the ways apps supported or stimulated discussions between parents and children about their online activities, it is closely related to previously defined active mediation in TOSS framework. However, in practice, active mediation features are more about \textit{active ways} in which restriction/monitoring can be implemented. This dimension portrayed the variation of such design details. This was done in two ways: first, through features that encouraged communication around the restriction and monitoring policies, and through coaching using discussion aides.

\begin{enumerate}
    \item (\textit{Low}) NO-MEANS: More than half of the apps (54\%) offered no means for children to negotiate screen rules with their parents, and their only choice was to accept and obey.
    \item (\textit{Medium}) SEND-MESSAGE: In comparison to the complete lack of communication support, these designs make it simple for children to send a message to their parents asking for permission to perform a particular restricted activity, or to grant an exception or extension to a particularly restrictive policy. These features have been identified in 18\% of our apps. 
    \item (\textit{Medium}) REWARD: Provide extrinsic reward incentives for children to earn additional screen time, app categories and content, or access to ordinarily restricted activities.
    \item (\textit{Medium}) BOTH DEVICE: We found one app that coach or prompt parents and children to spend time with each other, e.g., using a family time lock screen feature for both parents and children's phones to lock both devices out for a duration of time per day to encourage both to spend offline time without the distraction of a screen.
    \item (\textit{Medium}) APP-INFO: Provide specific information to help parents ascertain whether an app was suitable for their children's use, to understand potential risks, and other issues about the apps they might want to discuss.
    \item (\textit{Medium}) AGE-GROUP-INFO: Help parents to compare their children's online activities (usually screen time) in comparison with other kids of the same age group.
    \item (\textit{High}) CO-CONFIG: Provide explicit support that encourages parents and children to set screen rules together, which included interfaces designed to serve as boundary negotiating artifacts ~\cite{lee2007boundary} for a joint resolution of activity restrictions and monitoring policies.
    \item (\textit{High}) PARENTING-TIPS: Offer advice to parents, include strategies ranging from how to apply restriction/monitoring policies to their children, to how to talk about sensitive issues such as bullying, stranger danger, online pornography, and sexting.
\end{enumerate}

\subsection{Thematic Analysis of Reviews}

This section presents an overview of the primary themes pertaining to the reasons children and parents, respectively, liked or disliked feature designs provided by parental control apps, and how their perspectives varied across the axes of different ways in which features were realised.

\subsubsection{\colorbox{shadecolor!20}{\textit{Coarse Granularity}}}
We first start with perceptions around granularity - how different feature designs varied in granularity influenced users' perceptions of features. For \textit{coarse} granularity, we specifically refer to the features that are based on \textbf{ALL-OR-NOTHING} filtering. Parents had to either block all contents/shown with every detail (e.g., every video watched, website visited, the apps opened,even logging google search \& keyword queries.) about children’s online activities, or having no access at all, leaving no middle ground in between. Both parents and children, children in particular, expressed grievances on these features:

\noindent\textbf{Disliked being overly surveilled and restricted [C]} - The most common grievance of children regarding \textbf{ALL-OR-NOTHING}  was the view that these designs not only enabled but nudged parents to set up overly restrictive controls and excessive surveillance.  Kids expressed resentment at the extent of both the restrictions and their surveillance, and reflected on the effects lives, welfare, and activities:

\begin{small}
\begin{quote}
  This literally blocks everything, I can't even read e-books on my
  phone. 
\end{quote}
\end{small}

\begin{small}
\begin{quote}
  This is insane. Now they can see EVERYTHING! From my browsing history
to what apps I downloaded, even my texts! Worst app ever! 
\end{quote}
\end{small}

\noindent Kids reflected on a variety of secondary effects of excessive
restrictions and surveillance.  One child discussed that they felt
that, beyond violating their privacy, it was particularly wrong that
their parents' access to their messages would compromise their
friends' privacy as well:

\begin{small}
\begin{quote}
  they can eves drop on your convos and stuff that you dont want them
  to hear [...] not only is it a violation of my
  privacy that i didnt permit, but it is of friends too that parents
  dont know about
\end{quote}
\end{small}

\noindent The pervasiveness and constancy of surveillance made it feel to one
as if these apps enabled their parents to ``stalk'' them:

\begin{small}
\begin{quote}
    I hate this app my mom is like stalking my life!!
\end{quote}
\end{small}

\noindent Even when not restricted, the perception that their social communications were being surveilled by their parents had a chilling effect that  indirectly 
forced them to cease communicating and be cut off from their friends.

\begin{small}
\begin{quote}
  I can't talk with my friends anymore, everything will be recorded! 
\end{quote}
\end{small}

\noindent Several comments discussed longer-term effects such restrictions were having on their well-being. One view was that restrictions were directly and immediately harmful because they broke essential lines of social support.

\begin{small}
\begin{quote}
  This is stupid. absolutely awful. this will ruin people's lives. I
  had severe depression and the only thing keeping me from killing my
  self was my friend who I could only talk to online. Now I'm fine but
  if I lose contact with that friend I will most likely get my
  depression back. Horrible and stupid app. 
\end{quote}
\end{small}



\noindent Beyond cutting off lines of social support, restrictions were seen to
prevent kids from apps and activities that they normally used to cope
with boredom and isolation, further undermining well-being:

\begin{small}
\begin{quote}
 I can't play a lot of games, and I can't watch YouTube. I've sat in my room for weeks doing
 nothing and practically getting depressed because there is literally NOTHING I can do! 
\end{quote}
\end{small}



\noindent Beyond the social and emotional aspects of children's lives, a few comments 
connected these restrictions to developmental and educational harms-namely,
how such restrictions impeded learning by depriving them of experiences 
and learning opportunities:

\begin{small}
\begin{quote}
  The internet is where kids discover and learn new things. And by
  restricting it, you're denying them that ability.
\end{quote}
\end{small}



\noindent\textbf{We don't want to know everything [P]} - Interestingly, the most common comments regarding \textbf{ALL-OR-NOTHING} from parents were the complaints that the app is showing them \textit{too many} things while leaving them with \textit{too few} choices. Parents expressed resentment in terms of wanting to be able to do more tailored controls based on children's individual needs:

\begin{small}
\begin{quote}
 You either completely stop your kids from using their phone, or absolutely no rules. It's just so silly, don't they know kids these days need to do homework ONLINE?
\end{quote}
\end{small}

With regard to monitoring features, parents showed confusion and some reported lost in the vast amount of information given by the apps:

\begin{small}
\begin{quote}
 I don't want to know everything! What's the point of showing all these location info to me? I'm not a control freak! I just want to protect him from the crazy stuffs online, that's all!
\end{quote}
\end{small}

\noindent \textbf{We do want to know everything [P]} - On the other hand, some parents cherished being able to know every single detail of their children’s lives, and that led on to them expressing multiple opposing views around children's rights to privacy - whether children deserved any rights to privacy at all. One stated that since parents were paying for the phone, they should be able to set the rules:
\begin{small}
\begin{quote}
	Don’t listen to these spoiled children. I pay the bills. I control the phone. You want to control you pay the bills! Very very simple equation! It’s a shame how kids here actually can fathom the thought of ``rights'' while living in their parents' home. Who made that joke up?
\end{quote}
\end{small}

\noindent In some cases, parents also talked about how their duty as parents to keep kids safe outweighed any claim to rights:
\begin{small}
\begin{quote}
	When it comes to social media, kids don't need privacy. It’s not even about the child as much as it is about others preying on them. I would hope none of the negative reviews are from parents. You should monitor everything your children do. That is our job as parents.
\end{quote}
\end{small}

\noindent A more lenient version of this was the view that rights to autonomy/privacy should be an earned privilege, not a fundamental right:
\begin{small}
\begin{quote}
	If you want more trust privacy, prove you can handle it with good choices to show your parents you are trustworthy!
\end{quote}
\end{small}

\noindent Others viewed their children as being too young to make decisions for themselves until they become adults, regardless of their age:
\begin{small}
\begin{quote}
	It’s not that we don’t trust them, but studies show they can’t make decisions or assess risk like adults can until age 25 or so.
\end{quote}
\end{small}

\subsubsection{\colorbox{shadecolor!20}{\textit{High Granularity}}}
The other extreme along the granularity axis is the feature designs that are highly granular - for this, we specifically refer to the designs that allow parents the freedom to configure apps, websites, videos on a ``per item" basis (\textbf{CONTROL/MONITOR-PER-APP}). Parents were now able (sometimes even required) to refine the specific controls/monitoring they want.

\noindent \textbf{Lost in choices due to lack of support [P]} - As we observed earlier, many parents brought up how they'd like to be able to do more refined configurations when they mediate their children's online activities. However, when actually offered with these choices, several of the parents' comments expressed confusion and disappointment regarding these designs. One common complaint was that they simply don't have the time to go through each settings one by one, and they don't have the time to review each app to see if it needs to be banned: 

\begin{small}
\begin{quote}
	I'm a working mum with a full time job with three kids. Although I appreciate the app designers' efforts in letting us make the decision. It's just not practical for us working parents to go through all the apps one by one. 
\end{quote}
\end{small}

\noindent In some cases, parents talked about how frustrated they were as they felt they were indirectly accused as irresponsible parents, and they were ``nudged'' to go back to banning everything:

\begin{small}
\begin{quote}
	I don't see the point in letting us choose which videos for children to watch. One, I can't sit through everything they watch, I have a job. Two, 5-year-olds quickly get bored of the old ones and they want more. This so-called ``refinement'' just gave me two choices: either let them watch whatever they want (God knows what's on there), or bans the whole platform.
\end{quote}
\end{small}

\noindent In most cases, parents expressed their needs for supporting resources that help them make restriction/monitoring rules:
\begin{small}
\begin{quote}
	I really wish there's something like an app version of the tv age guide.
\end{quote}
\end{small}

\subsubsection{\colorbox{shadecolor!20}{\textit{Medium Granularity}}}
Feature designs of \textit{medium} granularity offer a middle ground for parents to mediate their children's online activities, without being too coarse or requires too much effort. Designs like this includes control/monitor based on app/website categories or age ratings; or control/monitor of contacts based on a parent-pre-approved contact; or apps reported and alerted parents only based on ``suspicious'' or ``dangerous'' activities, messages, or content.

\noindent\textbf{Protection, not punishment: achieving a successful middle ground [P]} - Parents were generally positive about the apps that offer feature designs of medium granularity, and they saw achieving a successful middle ground being essential for striking a balance between protection and respect for their children. This was seen not only as convenient for parents, but also supporting setting of boundaries in a more flexible manner. Similarly, both parents and children appreciated how some apps enabled different children, especially older ones, with different level of freedom: 
\begin{small}
\begin{quote}
    I love how this app allows us to reach balance. The app only alerts us when it detects something unusual, we can adjust the things he could access as he gets older.
\end{quote}
\end{small}

\noindent And when such balance was achieved, parents generally saw the parental control apps as effective at helping them support their primary goal, which was to keep their children safe. These parents felt such apps gave them ``peace of mind'':

\begin{small}
\begin{quote}
	I love this app. I have a 10 year old son that I just recently found out was doing inappropriate things on the Internet with his phone, like viewing porn [...] but with this app I Have some peace of mind. I have full control of what he downloads and views [...] all form [sic] my device. 
\end{quote}
\end{small}

\noindent\textbf{Reasonable safe zone [C]} - Similarly, when app designs managed to help strike a balance between protection and punishment, children regarded them as effective and reasonable at protecting them from dangers online:
\begin{small}
\begin{quote}
	Not too bad, I guess a bit boundary is necessary, at least I still have access to things I love. 
\end{quote}
\end{small}

\begin{small}
\begin{quote}
    This is just great. It allows age-appropriate control, so giving us more freedom as we get older. 
\end{quote}
\end{small}

\subsubsection{\colorbox{shadecolor!20}{\textit{Low Feedback/Transparency}}}
A second axis of feature design variation we looked at is feedback/transparency, and how users', especially children's perceptions varied across different designs along this axis. The designs came in various details along their ability of supporting children to have sufficient feedback and transparency, thus to allow them to learn and understand the screen rules, as well as learn more about their own activities online. For \textit{low} feedback/transparency, we refer to the designs that support very little or no means for children to become acknowledged of these things. Typical designs include apps provided no indication about the restrictions enforced on children's phones, leaving them with no idea of the things they could or could not do on their phones; apps did not acknowledge children about what information of them can be seen by parents; and apps only showed prompts at the moment when children were attempting to access content, however, they did not inform children with the screen rules in advance.

\noindent\textbf{Insecure, and not respected[C]} - Both children and parents, children in particular, expressed their dislike of feature designs that supported no feedback/transparency, which were sadly the most common feature designs. Rules without prior acknowledgment were sometimes presumed to be annoying by children. For example, both parents and children brought up that prior warnings on time remaining are important. Otherwise, children will feel upset due to the sudden cut-off. Children complained about how some app did not let them know what their screen rules were when such system was missing, and children found themselves in the position of having to work things out based on trial-and-error:
\begin{small}
\begin{quote}
	I constantly get these error messages with blank pages, is that part of the screen rules or just an error?
\end{quote}
\end{small}

In terms of monitoring, children felt insecure when not knowing what their parents can see about them:
\begin{small}
\begin{quote}
	Can they see all my texts too? That would be creepy.
\end{quote}
\end{small}

\subsubsection{\colorbox{shadecolor!20}{\textit{High Feedback/Transparency}}}

Feature designs that were of \textit{high} feedback/transparency includes designs not only provided children with means of viewing/inspecting the restriction policies but also offered a clear explanation on why the screen rules were made and why accessing restricted content might be bad for them. In terms of monitoring, features designs that support high feedback/transparency include the ones that provided children with feedback about their activities (allowing children to do self-monitoring), and offered children with detailed information about each app, including expert reviews and ratings and enabling children to decide what apps are good to use for themselves.

\noindent\textbf{Keeping them safe and productive [C]} - Both parents and children liked feature designs that supported children’s understanding of their online boundaries. In terms of restriction, both parents and children brought up how they liked apps offering children with clear screen rule pages so their children would be better informed instead of suffering from being turned down at the moment when they were trying to do some online activities. In particular, many kids pointed out how they liked being offered with explanations on why a website/app was blocked, which made them feel less confused and more respected as a consequence:

\begin{small}
\begin{quote}
    Not the best but definitely better than the previous one. Now I know why the websites are blocked. They give you reasons and things like that. I mean, I disagree with them all the time, but at least they tried to show some respect! 
\end{quote}
\end{small}

When children understood why the apps were for, they reported more positively about the apps. Several of the comments by children pointed out that the apps helped to keep them safe online from inappropriate things:
\begin{small}
 \begin{quote}
     I know this may sound crazy from the kids view,  but I love this now! It fits me and my phone perfectly, and my mom knows that I am safe on my phone without having to go to any other horrible apps.
 \end{quote}
\end{small}

\noindent Other children who suffered from addiction online appreciated how the apps dragged them out of that cycle:
\begin{small}
\begin{quote}
     This is AWESOME! I'm a kid so I got this app in order to keep myself in check on my screen time because I am an internet addict. I'm so much happier now! It makes it absolutely impossible to get around!
 \end{quote}
\end{small}

\noindent Beyond the positive impacts online, children also appreciated how the apps helped with their time offline. Some children commented that they were now able to become more productive and spend more time with their families:
\begin{small}
 \begin{quote}
     When my mom and dad put this on my phone and tablet at first I hated it. but then I realised with a limited amount of time I spent more time with my family and do actual work. I hope this inspires you to limit yourself with the amount of time you spend staring at useless junk. 
 \end{quote}
\end{small}

\noindent\textbf{Supporting them to make own decisions [C]} - In particular, children reported positively about designs that offer them with detailed information (including expert reviews and ratings) of each app. They talked about how they cherish being respected to make their own decisions on what is best for themselves:
\begin{small}
 \begin{quote}
    They tell you what others said about this app, but let you decide to use it or not – it’s my call.
 \end{quote}
\end{small}

\subsubsection{\colorbox{shadecolor!20}{\textit{No/Little Communications Support}}}
A third axis of feature design variation was \emph{parent-child communications support}. Specifically, we looked at how features were perceived differently due to the variation of communications support designs they came in with. For \textit{little} communications support, we refer to the designs that first, did not offer any explanation helping children to understand their screen rules, and also, offered none or very little means for children to negotiate screen rules with their parents, leaving them with the only choice - to accept and obey.

\noindent\textbf{Role of Parenting Apps: Unnecessary, Punishment, or Lazy Parenting [C]} -
When the purpose of these apps were not communicated with children, several of the children's comments expressed confusion or questioned the role these apps in keeping them safe. One child viewed that these apps were  fundamentally redundant because they were already old enough or competent enough to keep themselves safe:


\begin{small}
 \begin{quote}
   I'm a 4.0 student.  Who is able to manage her school work and
   screen time by herself thank you very much. I'm old enough to know
   what's good and bad, I can't change the settings and there no way to let my mom know that. 
 \end{quote}
\end{small}

\noindent Other kids lamented that these apps were unnecessary because kids could be told what to do and trusted to keep what they were told. In a sense, the use of apps to force represented a failure of trust.
 
\begin{small}
 \begin{quote}
    How bout you try and talk to them about your phone usage first, see if they'll make a change for you first, then go from there. Think twice before you destroy you kid's trust like this. I understand there's kids who need to get their head straight. But for those like me who are focused in school and a well-rounded kid, I feel all you need to do is talk to them.
 \end{quote}
\end{small}


\noindent The perception that parents were not trusting them made children question the point of these apps. In some cases, children regarded the use of these apps as purely for punishment.

\begin{small}
  \begin{quote}
    If you are a parent that wants this app I would reconsider getting this and punish your child a different way. Instead of us being punished, parents should be blamed for dictating their children to use this app.
 \end{quote}
\end{small}
 
\noindent Some children continued to criticise the bad parenting styles that these apps nudged their parents into. Apart from parents being overly-protective (as we previously reported), children also accused parents of being lazy and using the apps as substitutes for parenting. 
\begin{small}
 \begin{quote}
      Yes, it's good to be one step ahead, but having an app to do it for you? You might as well call child protective services if you're that lazy. 
 \end{quote}
\end{small}


\subsubsection{\colorbox{shadecolor!20}{\textit{High Communications Support}}}

Designs that were of \textit{high} communications support mainly came in two ways: first, through features that encouraged communications around the restriction and monitoring policies and through coaching using discussion aides. These includes designs such as apps enabling children to negotiate their rules with their parents, apps providing explicit support that encouraged parents and children to set screen rules together, and apps offering advice to parents - strategies ranging from how to apply restriction/monitoring policies to their children to how to talk about sensitive issues.

\noindent\textbf{Feeling of being respected [C]} - When children were given the chance to communicate and negotiate with their parents, they felt they were part of the decision and they generally respected the rules more. In particular, they loved the co-configuration designs that allowed them to sit down with their parents to reach mutual agreement on their boundaries:
\begin{small}
\begin{quote}
	My parents sat down with me to go through this ``setting rules together'' thing, and I can send a request to them whenever I felt the restrictions are unfair. Love it!
\end{quote}
\end{small}

\noindent\textbf{Supports, rather than enacts parenting[P]} -  Meanwhile, parents also reported their favour towards feature designs that supported or coached them on conversations with their children. And they found designs like parenting tips particularly useful. Some parents felt that they were supported by the app without being hijacked by it. 

\begin{small}
\begin{quote}
	This is a great tool. Easy to communicate. Easy to adjust. I use this tool to help, but in noway does it replace being a good parent watching over their child. It should not be the only arrow in your quiver, and you should not expect it to do your parenting for you.
\end{quote}
\end{small}

\section{Discussion}
\subsection{Connecting Feature Designs to Parental Mediation Theory}
Given the potential for both detrimental and beneficial effects of restriction and monitoring, and the likelihood that these strategies will remain important aspects of many parents' mediation strategies for the foreseeable future, can we strike a balance? Specifically, how might apps play a role in fostering a more responsible, considered, and appropriate use of these strategies? To answer these questions, we first reflect on whether some of our findings related to the feature design space described in Section \ref{subsection:threeaxes} might help. The three axes deal with very different aspects of parenting support and thus take each in turn.

Starting with the first axis, \emph{granularity} was associated with the degree and detail with which parents could control and monitor children's activities. This axis was directly related to the ``restrictive mediation” from parental mediation theory ~\cite{livingstone2008parental}. Beyenes et al.~\cite{beyens2017parent} found that restrictive mediation leads to increased conflicts between parents and children . Nathanson~\cite{nathanson2013media} argued that children might exhibit more disobedience and rebelliousness against parents under restrictive mediation . Ghosh et al.~\cite{10.1145/3173574.3173768} also found that teens who were victimized online and had peer problems were more likely to be controlled and monitored by their parents. These previous findings reinforce our analysis, such that a balance between restriction and children’s autonomy is much needed. The use and application of highly granular restrictions provides more flexibility for the parents but they can also be seen as overly demanding - parents need to be familiar with all the apps and websites their children could possibly access. On the other hand, the coarsest (all-or-nothing) controls were generally seen as less desirable because such apps denied children access to basic, risk-free, and unproblematic aspects of their devices that they deemed essential. The former takes a parent-centric view whilst permits limited inputs from children; and the latter is a close representation of an authoritarian parental approach ~\cite{darling1999parenting}, which has been deemed to decrease children and parents’ mutual trust and the development of children’s self-autonomy~\cite{10.1145/3173574.3173768, mensah2013influence} Various kinds of “medium granularity” identified in this research (e.g controlling by categories) were deemed most favoured by many parents because of the simplicity such granularities afford.  Such abstractions encourage slightly less parental involvement because of the way they homogenise the treatment of different apps in the same category; without requiring parents to be deeply involved with the particular risks, details or benefits of specific apps. 

With respect to \emph{transparency/feedback}, this axis dealt with the extent to which children remain informed of their own online activities and screen policies. By clearly establishing what was being restricted, monitored, and why, children could understand the boundaries of their parents' control so that they could negotiate their activities within and around such barriers. As such, this axis was most directly aligned with aspects of active \emph{self-regulation} ~\cite{fingerman2011handbook, kopp1982antecedents, moilanen2015bidirectional} but it's more than that - through empowering children with information and choices about their own behaviors online, as well as supporting resources such as information or expert ratings on particular apps, children were given the opportunity to make more responsible decisions for themselves  - one of the key outcomes from an active mediation approach ~\cite{shin2017does}. Less feedback and transparency could be seen as \emph{more privacy-invasive and autonomy restrictive}, making it less clear or visible when and what parents would be watching. Much like Bentham's panopticon ~\cite{strub1989theory}, less transparency can be seen as undermining children's rights by creating a perpetual spectre of possible observation.  

Regarding \emph{communications support}, this axis was most closely related to the ``active mediation” from parental mediation theory, in which communication is the core ~\cite{livingstone2008parental, lee2007children, nikken2014developing}. Previous studies found that active mediation was associated with lower levels of aggression children ~\cite{collier2016does}, and that active mediation may reduce online risks without hindering children's online opportunities ~\cite{duerager2012can, wisniewski2015preventative}. Kuppens et al.~\cite{kuppens2019parenting} found that children who experienced frequent discussions about online activities with parents were more likely to regard control/monitoring as legitimately in their best interests . While being closely related to active mediation, this axis was more than that. It dealt with providing supports for parent-child interactions in several forms; one was parental informational support, which comprised both \emph{general parenting advice} pertaining to ways to talk to children about their online activities and risks, and \emph{specific information} about particular apps and activities. Other support includes encouraging the co-setting of restrictions and monitoring practices, allowing children to be involved in the configuration of such activities, and can have significant benefits in terms of how they view such activities as a result ~\cite{kuppens2019parenting}.

We contend that granular restrictions and monitoring support active/authoritative parenting \mbox{\cite{darling1999parenting}}  by increasing the flexibility and means by which parents can remain in-the-loop of, and control over, children's activities. However, we argue that our finding shows that this is only true for apps which also had a high degree of \emph{transparency/feedback} (axis 2), as well as high \mbox{\emph{communications support}} (axis 3). Such capabilities are essential for enabling children to understand restrictions, monitored activities, and goals (transparency/feedback), providing a means of understanding of the context and purpose of these restrictions, and a means of digitally facilitating discussion, negotiation, or relaxation of such policies (communications support). Without such feedback, monitoring and restrictions are most directly aligned to \mbox{\emph{authoritarian parenting}} \mbox{\cite{ford1985factors, darling1999parenting}} by enabling precise control and detailed monitoring without associated feedback, transparency, or support for negotiation or communications. Parental mediation theory categorised parental mediation strategies into distinct types, namely restrictive, monitoring and active ~\cite{livingstone2008parental}. However, a new view of parental mediation as a continuous, dynamic process is supported by recent empirical and theoretical literature, including work by Symonds et al.~\cite{symons2017parental}, who proposed that parental mediation be viewed as a dynamic integrative process that brings strategies into the daily interactions between parents and children. Other studies that support this view similarly found that parents tended not to use restrictive, monitoring, or active mediation strategies independently, but in ways that supported each other~\cite{beyens2017parent, meeus2018managing}, while Chen et al.~\cite{chen2016active} suggested that mediation such as restriction and monitoring were most effective when they were implemented in an active way. Our findings reinforced this new look of the parental mediation approaches: restrictive and active mediations are not in and of themselves good or bad; rather, their effectiveness may be highly dependent on the style in which parents apply these strategies with their children ~\cite{valkenburg2013developing, chen2016active}. The feature designs along three axes (granularity, transparency/feedback, communications support) align perfectly with these arguments - instead of considering features as separate categories, the axes of variation reflect the designs on how strategic elements can be combined, and implemented in a lesser or greater extent - overlapping and reinforcing each other, in an overall dynamic process.

\subsection{Revisiting the TOSS Framework}

\begin{figure}[h]
    \centering
    \includegraphics[width=1\textwidth]{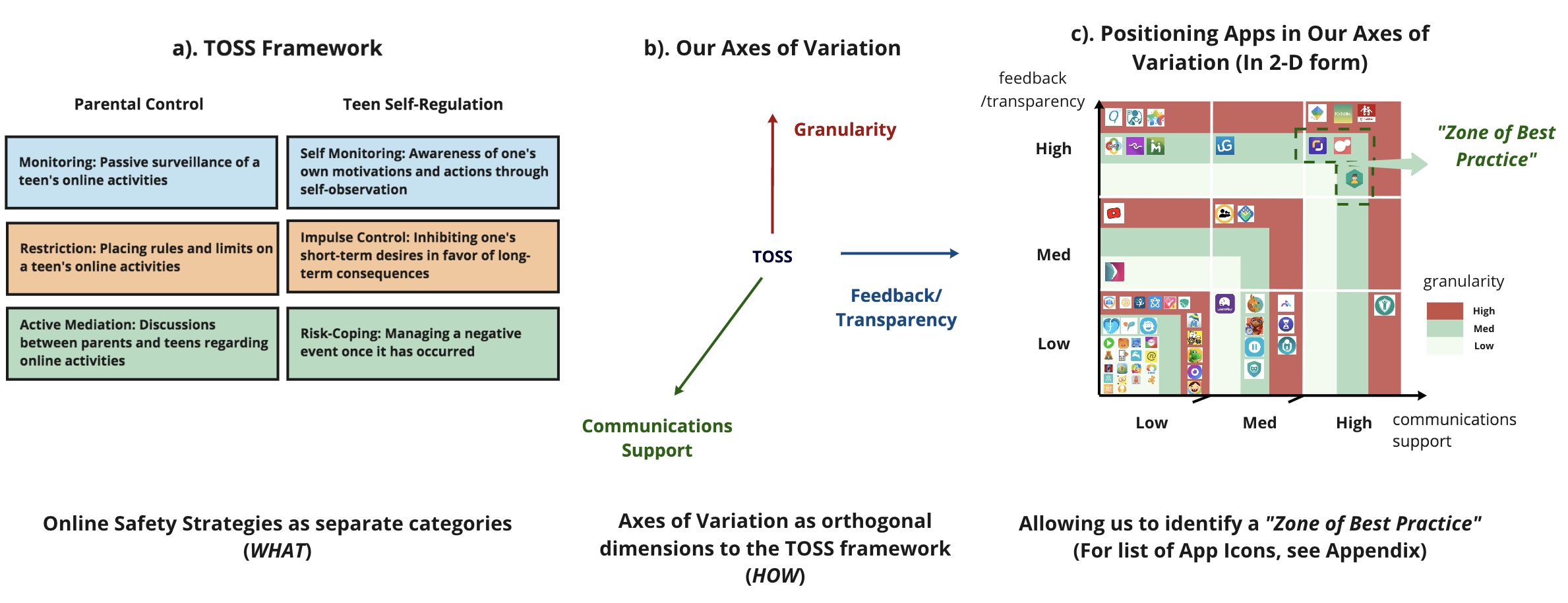}
    \caption{Revisiting the TOSS Framework. a). The TOSS framework looks at \textit{what} features are supported b). Our Axes of Variation complement the TOSS framework by adding orthogonal dimensions of \textit{how} features were implemented. c). As a result, our three axes of variation allow us to identify a \textit{Zone of Best Practice}, this also shows that most of the apps on the current market fall outside of this zone.}
    \label{fig:rev}
\end{figure}

Our analysis offered empirical evidence that suggests that the ways in which features are designed influence users' perceptions of features in parental control apps.  In particular, we found that parents and children's perceptions of features varied significantly across the three axes designs variation. For example, for a single restriction feature such as APP-BLOCK, when implemented differently, such with features for supporting parents-child negotiation, were viewed more positively than when apps offered no such means at all. 

Instead of viewing app features as belonging to separate independent mediation strategies as TOSS does, we prefer to view such features as playing variable roles in mediation strategies articulated by many factors including their design. We feel that our axes of variation can be seen as an extension to the TOSS framework, orthogonal to its categorisation of features by strategy, that provides a view to the ways such features should be designed.  In particular, by looking at the areas along each of the axes that correspond to the most positive comments, we can identify a  \textit{Zone of Best Practice}, corresponding to the regions of the feature design space viewed most positively by parents and children. For \textit{granularity}, the designs that were of medium level of granularity were perceived as helping to strike a balance between protecting children online, while leaving them with enough space to learn and explore online - such designs include configurations on categorical level or age ratings, managing contacts based on a parent-pre-approved list, or only alerting parents when ``suspicious'' or ``dangerous'' activities were identified. For \textit{feedback/transparency}, the most welcomed designs were those that provided support for children to remain informed of their screen rules, their own activities online, and support them to make decisions for themselves online - good design practices include panels that show children explicit information on their own online activities, pop ups that explain why certain screen rules were made for them, and info panels that offer children with detailed information of each app, thus enabling themselves to decide what apps to use. For \textit{communications support}, the most welcomed designs were those that encourage negotiation between parents and children, and support parents with information and resources to start those conversations - e.g. co-configuration panels that encourage parents and children to set screen rules together, pop up parenting tips that offer advice to parents. Unfortunately, our analysis also found that most existing apps fell outside this zone for many of their features.

\subsection{Opportunities for more autonomy-supportive parental mediation}

One of the major problems with restriction and monitoring is the extent to which these strategies reduce children's autonomy and violate their privacy.  How might apps be made to be more autonomy-supportive, especially in the long term? Our view is that the inherent problem with the parental control apps we reviewed is that they were focused exclusively on effective restriction and monitoring, instead of digital parenting. Instead of framing restrictions and monitoring as protective measures to be used in perpetuity, for instance, a digital parenting app might not only include support for other TOSS strategies (such as impulse control), but also re-frame restriction and monitoring as a means of skills scaffolding \cite{hammond2012effects, bibok2009parental}.  Starting with the most supportive scaffolding (maximum restrictions and monitoring) for the very young and most vulnerable, such apps could then align the gradual removal of restrictions and monitoring measures with milestones as children get older, or demonstrate the ability to recognise and cope with online risks. Then, as their needs for autonomy and privacy needs change, typically peaking by mid-teenage years~\cite{arnett2000emerging}, such restrictions should probably reach a minimum at such an age. However, if children require special support such scaffolds could be seen as flexible as needed. Such a re-framing would address several of the children's comments we found, especially relating to children seeing restrictions as pointless and infantilising, and supporting them as they grow older, so as to not be seen to hold them back.

Another re-framing that might provide protection against overuse of restriction and monitoring is to frame such measures as options of last resort, after other more autonomy-confirming strategies fail. Primary autonomy-confirming strategies for older children might instead include, for example, simple verbal agreements about limits on online activity which require children to take responsibility for their actions.  Such a shift would be in line with the literature on autonomy-supportive parental mediation, which requires a  ``convincing rationale for rule-making''~\cite{haddon2015children}.  We propose that similarly, requiring a convincing rationale to justify the use of restriction and monitoring features might ensure that they are used only as necessary.

\section{Limitations and Future Work}

There are several important limitations of this work; the first pertains to the app analysis; our sample of parental control apps were taken only from the Android UK Google Play store, and thus may vary significantly by region.  Due to a feasibility constraints we were only able to analyse 58 of these apps, out of the massive growing pool available.  The second most significant limitation is that we based our analysis on reviews available in the wild.  There are obvious limitations of trying to use reviews as ``vessels of truth''; namely, they could be fabricated by bots, or, even if genuine, could represent a skewed non-representative sample of particularly angry or passionate people.  We attempted to mitigate some of these possibilities several ways, one was by using a process of filtering reviews to carefully select only those discussing specific aspects of apps. Moreover, we avoided deriving any conclusions through quantitative measures which would be distorted; instead, we used reviews only to collate reasoned perspectives about features.

Nonetheless caution should still be exercised using any app reviews for deriving any sort of definitive conclusions about causal relationships, for instance, relating app use to children's welfare or safety. Our contributions thus should be interpreted as a high-level understanding of the design space of features of parental control apps, with evidence that might suggest regions most promising for further exploration. Our future work would involve actually working with parents and children to get more direct inputs.  We hope our contributions could guide others navigate the space of parental control apps, for running field studies of the effectiveness of specific feature designs.  

Apart from that, this work adopted a ground-up approach as well as being guided by the TOSS framework, instead of a value-driven approach. Therefore, we did not take a full consideration of the different cultural and parenting styles intrinsically embedded in different cultures and societal values, which would benefit from future work.

\section{Conclusion}


Parental control apps are being increasingly seen as the answer to online safety concerns parents have for their children~\cite{ofcom2019}. However, increasing evidence suggests that such apps may be bringing with them new kinds of harms associated with excessive restrictions and privacy invasion~\cite{wisniewski2017parental, ghosh2018safety}. This paper is the first to contribute an understanding of the design space of the mediation features of current parental control apps. Then, to identify how such variation might articulate the use of these apps in practice, we analysed reviews corresponding to features at various points along each of the axes.  Our findings suggest that apps providing the most transparency and feedback, the greatest support for parent-child communications, and flexible categories are most likely to be perceived by both parents and children in positive ways instead as punitive and detrimental.  Contextualising our findings in parental control theory, we derive at least two approaches that more autonomy-supportive parental control apps might be achieved.  The first, by simply designing app features to fall along each of the three axes within our identified ``zone of best practice''--designs  perceived most positively by both parents and children.  Second, to complement and pair restriction and monitoring features in an age and skill-appropriate manner. This includes gradually allowing children to move from strictest monitoring and restriction features to gain more autonomy, as they develop necessary skills associated with recognising and mitigating online risks.

\section{Acknowledgments}
This work was supported by the ReEnTrust project from the Engineering and Physical Sciences Research Council (grant number EP/R033633/1) and Oxford Univeristy COVID-19 Rebuilding Research Momentum Fund (0010067).

\appendix

\section{Appendix}

\begin{figure}[H]
    \centering
    \includegraphics[width=1\textwidth]{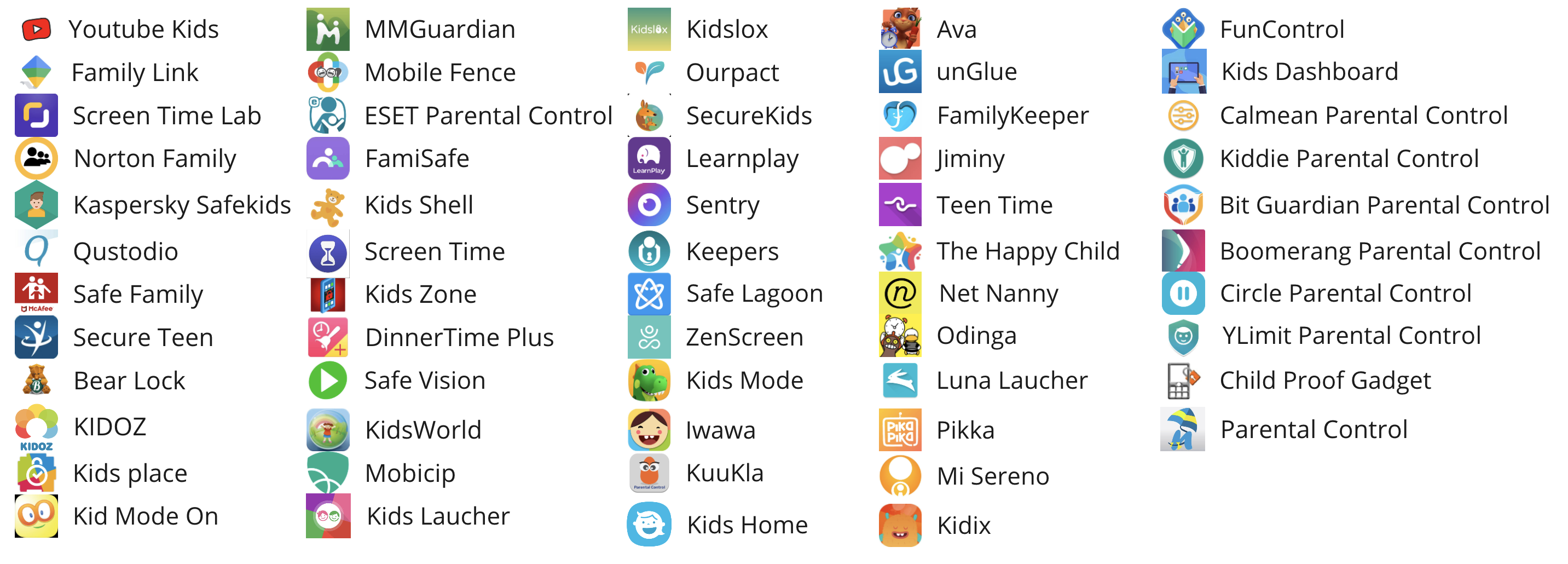}
    \caption{App List with Icons}
\end{figure}

\begin{figure}[H]
    \centering
    \includegraphics[width=1\textwidth]{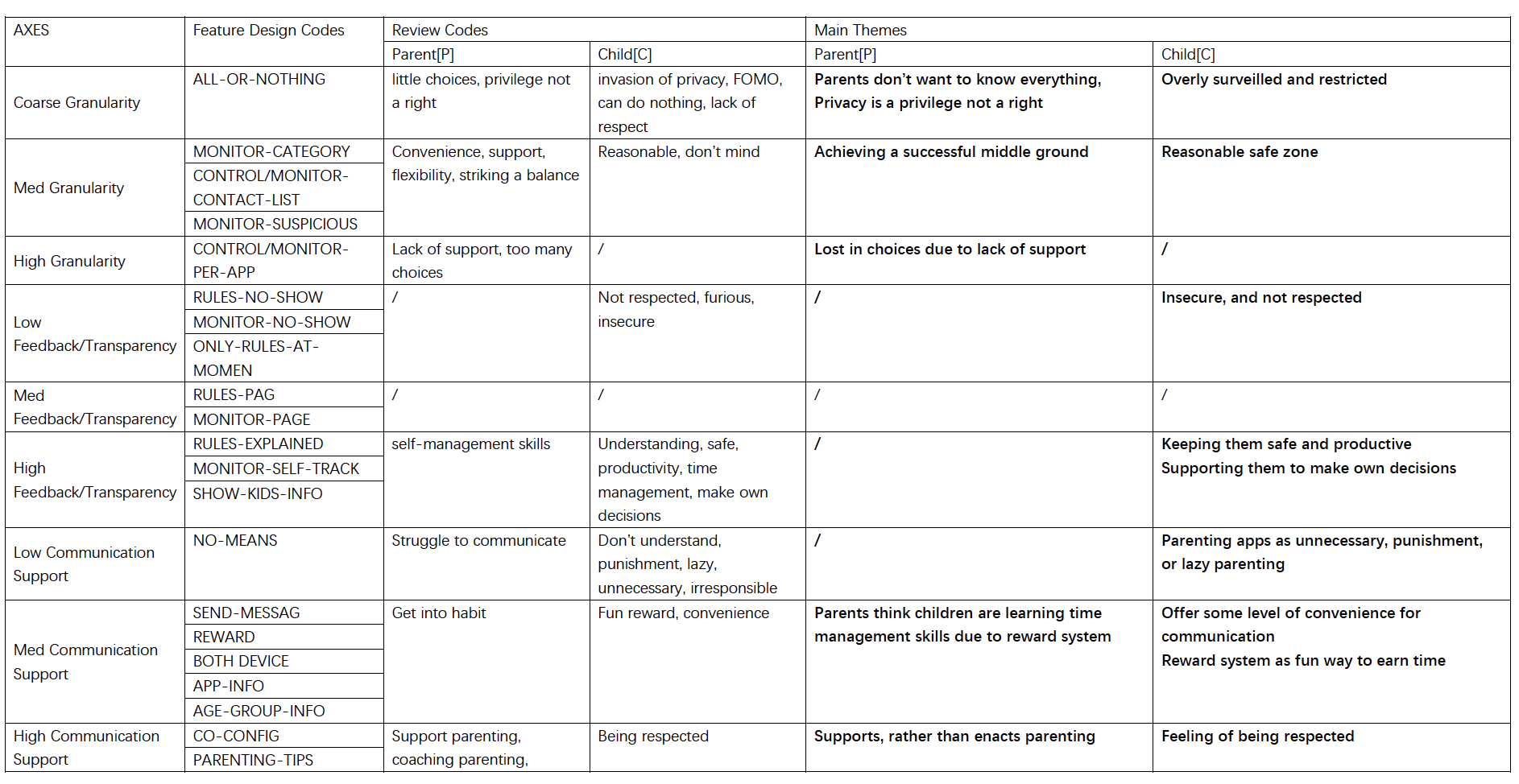}
    \caption{Final Codebook for Reviews. Reviews were first organised using Feature Designs Codebook we previously compiled in Section \ref{subsubsection:methodFeatureDesign}. Then, we used a grounded, thematic approach~\cite{wildemuth2016applications} to identify themes about why a parent or a child liked or disliked that feature design.}
\end{figure}


\begin{thebibliography}{78}


\ifx \showCODEN    \undefined \def \showCODEN     #1{\unskip}     \fi
\ifx \showDOI      \undefined \def \showDOI       #1{#1}\fi
\ifx \showISBNx    \undefined \def \showISBNx     #1{\unskip}     \fi
\ifx \showISBNxiii \undefined \def \showISBNxiii  #1{\unskip}     \fi
\ifx \showISSN     \undefined \def \showISSN      #1{\unskip}     \fi
\ifx \showLCCN     \undefined \def \showLCCN      #1{\unskip}     \fi
\ifx \shownote     \undefined \def \shownote      #1{#1}          \fi
\ifx \showarticletitle \undefined \def \showarticletitle #1{#1}   \fi
\ifx \showURL      \undefined \def \showURL       {\relax}        \fi
\providecommand\bibfield[2]{#2}
\providecommand\bibinfo[2]{#2}
\providecommand\natexlab[1]{#1}
\providecommand\showeprint[2][]{arXiv:#2}

\bibitem[\protect\citeauthoryear{??}{mrf}{2019}]%
        {mrfr}
 \bibinfo{year}{2019}\natexlab{}.
\newblock \showarticletitle{Global Parental Control Software Market Research
  Report}.
\newblock  (\bibinfo{year}{2019}).
\newblock
\urldef\tempurl%
\url{https://www.marketresearchfuture.com/reports/parental-control-software-market-4551}
\showURL{%
\tempurl}


\bibitem[\protect\citeauthoryear{??}{gps}{2020}]%
        {gps}
 \bibinfo{year}{2020}\natexlab{}.
\newblock \showarticletitle{{google-play-scraper 0.1.1}}.
\newblock  (\bibinfo{year}{2020}).
\newblock


\bibitem[\protect\citeauthoryear{5Rights}{5Rights}{2018}]%
        {disruptedchildhood}
\bibfield{author}{\bibinfo{person}{5Rights}.} \bibinfo{year}{2018}\natexlab{}.
\newblock \bibinfo{title}{Disrupted Childhood: The Cost of Persuasive Design}.
\newblock
\newblock
\urldef\tempurl%
\url{https://5rightsframework.com/static/5Rights-Disrupted-Childhood.pdf}
\showURL{%
\tempurl}


\bibitem[\protect\citeauthoryear{Alelyani, Ghosh, Moralez, Guha, and
  Wisniewski}{Alelyani et~al\mbox{.}}{2019}]%
        {alelyani2019examining}
\bibfield{author}{\bibinfo{person}{Turki Alelyani}, \bibinfo{person}{Arup~Kumar
  Ghosh}, \bibinfo{person}{Larry Moralez}, \bibinfo{person}{Shion Guha}, {and}
  \bibinfo{person}{Pamela Wisniewski}.} \bibinfo{year}{2019}\natexlab{}.
\newblock \showarticletitle{Examining Parent Versus Child Reviews of Parental
  Control Apps on Google Play}. In \bibinfo{booktitle}{\emph{Int'l
  Conf. on Human-Computer Interaction}}. Springer, \bibinfo{pages}{3--21}.
\newblock


\bibitem[\protect\citeauthoryear{Ali, Blades, Oates, and Blumberg}{Ali
  et~al\mbox{.}}{2009}]%
        {ali2009young}
\bibfield{author}{\bibinfo{person}{Moondore Ali}, \bibinfo{person}{Mark
  Blades}, \bibinfo{person}{Caroline Oates}, {and} \bibinfo{person}{Fran
  Blumberg}.} \bibinfo{year}{2009}\natexlab{}.
\newblock \showarticletitle{Young children's ability to recognize
  advertisements in web page designs}.
\newblock \bibinfo{journal}{\emph{British Journal of Developmental Psychology}}
  \bibinfo{volume}{27}, \bibinfo{number}{1} (\bibinfo{year}{2009}),
  \bibinfo{pages}{71--83}.
\newblock


\bibitem[\protect\citeauthoryear{Arnett}{Arnett}{2000}]%
        {arnett2000emerging}
\bibfield{author}{\bibinfo{person}{Jeffrey~Jensen Arnett}.}
  \bibinfo{year}{2000}\natexlab{}.
\newblock \showarticletitle{Emerging adulthood: A theory of development from
  the late teens through the twenties.}
\newblock \bibinfo{journal}{\emph{American psychologist}} \bibinfo{volume}{55},
  \bibinfo{number}{5} (\bibinfo{year}{2000}), \bibinfo{pages}{469}.
\newblock


\bibitem[\protect\citeauthoryear{Bandura}{Bandura}{[n.d.]}]%
        {bandura1977social}
\bibfield{author}{\bibinfo{person}{Albert Bandura}.}
  \bibinfo{year}{[n.d.]}\natexlab{}.
\newblock \bibinfo{booktitle}{\emph{Social learning theory}}.
  Vol.~\bibinfo{volume}{1}.
\newblock


\bibitem[\protect\citeauthoryear{Beyens and Beullens}{Beyens and
  Beullens}{2017}]%
        {beyens2017parent}
\bibfield{author}{\bibinfo{person}{Ine Beyens} {and} \bibinfo{person}{Kathleen
  Beullens}.} \bibinfo{year}{2017}\natexlab{}.
\newblock \showarticletitle{Parent--child conflict about children’s tablet
  use: The role of parental mediation}.
\newblock \bibinfo{journal}{\emph{New Media \& Society}} \bibinfo{volume}{19},
  \bibinfo{number}{12} (\bibinfo{year}{2017}), \bibinfo{pages}{2075--2093}.
\newblock


\bibitem[\protect\citeauthoryear{Bibok, Carpendale, and M{\"u}ller}{Bibok
  et~al\mbox{.}}{2009}]%
        {bibok2009parental}
\bibfield{author}{\bibinfo{person}{Maximilian~B Bibok},
  \bibinfo{person}{Jeremy~IM Carpendale}, {and} \bibinfo{person}{Ulrich
  M{\"u}ller}.} \bibinfo{year}{2009}\natexlab{}.
\newblock \showarticletitle{Parental scaffolding and the development of
  executive function}.
\newblock \bibinfo{journal}{\emph{New directions for child and adolescent
  development}} \bibinfo{volume}{2009}, \bibinfo{number}{123}
  (\bibinfo{year}{2009}), \bibinfo{pages}{17--34}.
\newblock


\bibitem[\protect\citeauthoryear{Bird, Klein, and Loper}{Bird
  et~al\mbox{.}}{2009}]%
        {bird2009natural}
\bibfield{author}{\bibinfo{person}{Steven Bird}, \bibinfo{person}{Ewan Klein},
  {and} \bibinfo{person}{Edward Loper}.} \bibinfo{year}{2009}\natexlab{}.
\newblock \bibinfo{booktitle}{\emph{Natural language processing with Python:
  analyzing text with the natural language toolkit}}.
\newblock \bibinfo{publisher}{" O'Reilly Media, Inc."}.
\newblock


\bibitem[\protect\citeauthoryear{Brian~O’Neill}{Brian~O’Neill}{2018}]%
        {policy20218}
\bibfield{author}{\bibinfo{person}{Thuy~Dinh Brian~O’Neill}.}
  \bibinfo{year}{2018}\natexlab{}.
\newblock \bibinfo{title}{The Better Internet for Kids Policy Map}.
\newblock
\newblock
\urldef\tempurl%
\url{https://www.betterinternetforkids.eu/documents/167024/2637346/BIK+Map+report+-+Final+-+March+2018/a858ae53-971f-4dce-829c-5a02af9287f7}
\showURL{%
\tempurl}


\bibitem[\protect\citeauthoryear{Briones, Madden, and Janoske}{Briones
  et~al\mbox{.}}{2013}]%
        {briones2013kony}
\bibfield{author}{\bibinfo{person}{Rowena Briones}, \bibinfo{person}{Stephanie
  Madden}, {and} \bibinfo{person}{Melissa Janoske}.}
  \bibinfo{year}{2013}\natexlab{}.
\newblock \showarticletitle{Kony 2012: Invisible children and the challenges of
  social media campaigning and digital activism}.
\newblock \bibinfo{journal}{\emph{Journal of Current Issues in Media and
  Telecommunications ISSN}}  \bibinfo{volume}{1935} (\bibinfo{year}{2013}),
  \bibinfo{pages}{3588}.
\newblock


\bibitem[\protect\citeauthoryear{Chaiklin et~al\mbox{.}}{Chaiklin
  et~al\mbox{.}}{2003}]%
        {chaiklin2003zone}
\bibfield{author}{\bibinfo{person}{Seth Chaiklin} {et~al\mbox{.}}}
  \bibinfo{year}{2003}\natexlab{}.
\newblock \showarticletitle{The zone of proximal development in Vygotsky’s
  analysis of learning and instruction}.
\newblock \bibinfo{journal}{\emph{Vygotsky’s educational theory in cultural
  context}} \bibinfo{volume}{1}, \bibinfo{number}{2} (\bibinfo{year}{2003}),
  \bibinfo{pages}{39--64}.
\newblock


\bibitem[\protect\citeauthoryear{Chen and Chng}{Chen and Chng}{2016}]%
        {chen2016active}
\bibfield{author}{\bibinfo{person}{Vivian Hsueh~Hua Chen} {and}
  \bibinfo{person}{Grace~S Chng}.} \bibinfo{year}{2016}\natexlab{}.
\newblock \showarticletitle{Active and restrictive parental mediation over
  time: Effects on youths’ self-regulatory competencies and impulsivity}.
\newblock \bibinfo{journal}{\emph{Computers \& Education}}
  \bibinfo{volume}{98} (\bibinfo{year}{2016}), \bibinfo{pages}{206--212}.
\newblock


\bibitem[\protect\citeauthoryear{Clark}{Clark}{2011}]%
        {clark2011parental}
\bibfield{author}{\bibinfo{person}{Lynn~Schofield Clark}.}
  \bibinfo{year}{2011}\natexlab{}.
\newblock \showarticletitle{Parental mediation theory for the digital age}.
\newblock \bibinfo{journal}{\emph{Communication theory}} \bibinfo{volume}{21},
  \bibinfo{number}{4} (\bibinfo{year}{2011}), \bibinfo{pages}{323--343}.
\newblock


\bibitem[\protect\citeauthoryear{Collier, Coyne, Rasmussen, Hawkins,
  Padilla-Walker, Erickson, and Memmott-Elison}{Collier et~al\mbox{.}}{2016}]%
        {collier2016does}
\bibfield{author}{\bibinfo{person}{Kevin~M Collier}, \bibinfo{person}{Sarah~M
  Coyne}, \bibinfo{person}{Eric~E Rasmussen}, \bibinfo{person}{Alan~J Hawkins},
  \bibinfo{person}{Laura~M Padilla-Walker}, \bibinfo{person}{Sage~E Erickson},
  {and} \bibinfo{person}{Madison~K Memmott-Elison}.}
  \bibinfo{year}{2016}\natexlab{}.
\newblock \showarticletitle{Does parental mediation of media influence child
  outcomes? A meta-analysis on media time, aggression, substance use, and
  sexual behavior.}
\newblock \bibinfo{journal}{\emph{Developmental psychology}}
  \bibinfo{volume}{52}, \bibinfo{number}{5} (\bibinfo{year}{2016}),
  \bibinfo{pages}{798}.
\newblock


\bibitem[\protect\citeauthoryear{Darling}{Darling}{1999}]%
        {darling1999parenting}
\bibfield{author}{\bibinfo{person}{Nancy Darling}.}
  \bibinfo{year}{1999}\natexlab{}.
\newblock \bibinfo{booktitle}{\emph{Parenting style and its correlates}}.
\newblock \bibinfo{publisher}{ERIC Clearinghouse on Elementary and Early
  Childhood Education, University~…}.
\newblock


\bibitem[\protect\citeauthoryear{Duerager and Livingstone}{Duerager and
  Livingstone}{2012}]%
        {duerager2012can}
\bibfield{author}{\bibinfo{person}{Andrea Duerager} {and}
  \bibinfo{person}{Sonia Livingstone}.} \bibinfo{year}{2012}\natexlab{}.
\newblock \showarticletitle{How can parents support children’s internet
  safety?}
\newblock  (\bibinfo{year}{2012}).
\newblock


\bibitem[\protect\citeauthoryear{Feal, Calciati, Vallina-Rodriguez, Troncoso,
  and Gorla}{Feal et~al\mbox{.}}{2020}]%
        {feal2020angel}
\bibfield{author}{\bibinfo{person}{{\'A}lvaro Feal}, \bibinfo{person}{Paolo
  Calciati}, \bibinfo{person}{Narseo Vallina-Rodriguez},
  \bibinfo{person}{Carmela Troncoso}, {and} \bibinfo{person}{Alessandra
  Gorla}.} \bibinfo{year}{2020}\natexlab{}.
\newblock \showarticletitle{Angel or Devil? A Privacy Study of Mobile Parental
  Control Apps}.
\newblock \bibinfo{journal}{\emph{Proceedings on Privacy Enhancing
  Technologies}} \bibinfo{volume}{2020}, \bibinfo{number}{2}
  (\bibinfo{year}{2020}), \bibinfo{pages}{314--335}.
\newblock


\bibitem[\protect\citeauthoryear{Fikkers, Piotrowski, and Valkenburg}{Fikkers
  et~al\mbox{.}}{2017}]%
        {fikkers2017matter}
\bibfield{author}{\bibinfo{person}{Karin~M Fikkers},
  \bibinfo{person}{Jessica~Taylor Piotrowski}, {and} \bibinfo{person}{Patti~M
  Valkenburg}.} \bibinfo{year}{2017}\natexlab{}.
\newblock \showarticletitle{A matter of style? Exploring the effects of
  parental mediation styles on early adolescents’ media violence exposure and
  aggression}.
\newblock \bibinfo{journal}{\emph{Computers in Human Behavior}}
  \bibinfo{volume}{70} (\bibinfo{year}{2017}), \bibinfo{pages}{407--415}.
\newblock


\bibitem[\protect\citeauthoryear{Fingerman, Berg, Smith, and
  Antonucci}{Fingerman et~al\mbox{.}}{2011}]%
        {fingerman2011handbook}
\bibfield{author}{\bibinfo{person}{Karen~L Fingerman}, \bibinfo{person}{Cynthia
  Berg}, \bibinfo{person}{Jacqui Smith}, {and} \bibinfo{person}{Toni~C
  Antonucci}.} \bibinfo{year}{2011}\natexlab{}.
\newblock \bibinfo{booktitle}{\emph{Handbook of life-span development}}.
\newblock \bibinfo{publisher}{Springer Publishing Company}.
\newblock


\bibitem[\protect\citeauthoryear{Ford, Massey, and Hyde}{Ford
  et~al\mbox{.}}{1985}]%
        {ford1985factors}
\bibfield{author}{\bibinfo{person}{Denyce~S Ford}, \bibinfo{person}{Kay~K
  Massey}, {and} \bibinfo{person}{David Hyde}.}
  \bibinfo{year}{1985}\natexlab{}.
\newblock \showarticletitle{Factors related to authoritarian versus
  nonauthoritarian attitudes toward parenting among college students}.
\newblock \bibinfo{journal}{\emph{Health Education}} \bibinfo{volume}{16},
  \bibinfo{number}{6} (\bibinfo{year}{1985}), \bibinfo{pages}{26--28}.
\newblock


\bibitem[\protect\citeauthoryear{Ghosh, Badillo-Urquiola, Guha, LaViola~Jr, and
  Wisniewski}{Ghosh et~al\mbox{.}}{2018a}]%
        {ghosh2018safety}
\bibfield{author}{\bibinfo{person}{Arup~Kumar Ghosh}, \bibinfo{person}{Karla
  Badillo-Urquiola}, \bibinfo{person}{Shion Guha}, \bibinfo{person}{Joseph~J
  LaViola~Jr}, {and} \bibinfo{person}{Pamela~J Wisniewski}.}
  \bibinfo{year}{2018}\natexlab{a}.
\newblock \showarticletitle{Safety vs. surveillance: what children have to say
  about mobile apps for parental control}. In
  \bibinfo{booktitle}{\emph{Proceedings of the 2018 CHI Conf. on Human
  Factors in Computing Systems}}. \bibinfo{pages}{1--14}.
\newblock


\bibitem[\protect\citeauthoryear{Ghosh, Badillo-Urquiola, Rosson, Xu, Carroll,
  and Wisniewski}{Ghosh et~al\mbox{.}}{2018b}]%
        {10.1145/3173574.3173768}
\bibfield{author}{\bibinfo{person}{Arup~Kumar Ghosh}, \bibinfo{person}{Karla
  Badillo-Urquiola}, \bibinfo{person}{Mary~Beth Rosson}, \bibinfo{person}{Heng
  Xu}, \bibinfo{person}{John~M. Carroll}, {and} \bibinfo{person}{Pamela~J.
  Wisniewski}.} \bibinfo{year}{2018}\natexlab{b}.
\newblock \showarticletitle{A Matter of Control or Safety? Examining Parental
  Use of Technical Monitoring Apps on Teens' Mobile Devices}. In
  \bibinfo{booktitle}{\emph{Proceedings of the 2018 CHI Conf. on Human
  Factors in Computing Systems}} (Montreal QC, Canada)
  \emph{(\bibinfo{series}{CHI '18})}. \bibinfo{publisher}{Association for
  Computing Machinery}, \bibinfo{address}{NY, USA},
  \bibinfo{pages}{1–14}.
\newblock
\showISBNx{9781450356206}
\urldef\tempurl%
\url{https://doi.org/10.1145/3173574.3173768}
\showDOI{\tempurl}


\bibitem[\protect\citeauthoryear{Ghosh, Hughes, and Wisniewski}{Ghosh
  et~al\mbox{.}}{2020}]%
        {ghosh2020circle}
\bibfield{author}{\bibinfo{person}{Arup~Kumar Ghosh},
  \bibinfo{person}{Charles~E Hughes}, {and} \bibinfo{person}{Pamela~J
  Wisniewski}.} \bibinfo{year}{2020}\natexlab{}.
\newblock \showarticletitle{Circle of Trust: A New Approach to Mobile Online
  Safety for Families}. In \bibinfo{booktitle}{\emph{Proceedings of the 2020
  CHI Conf.e on Human Factors in Computing Systems}}.
  \bibinfo{pages}{1--14}.
\newblock


\bibitem[\protect\citeauthoryear{Government}{Government}{2019}]%
        {hm2019online}
\bibfield{author}{\bibinfo{person}{HM Government}.}
  \bibinfo{year}{2019}\natexlab{}.
\newblock \bibinfo{title}{Online Harms White Paper}.
\newblock
\newblock


\bibitem[\protect\citeauthoryear{Guzman and Maalej}{Guzman and Maalej}{2014}]%
        {guzman2014users}
\bibfield{author}{\bibinfo{person}{Emitza Guzman} {and} \bibinfo{person}{Walid
  Maalej}.} \bibinfo{year}{2014}\natexlab{}.
\newblock \showarticletitle{How do users like this feature? a fine grained
  sentiment analysis of app reviews}. In \bibinfo{booktitle}{\emph{2014 IEEE
  22nd 'Int'l requirements engineering conf. (RE)}}. IEEE,
  \bibinfo{pages}{153--162}.
\newblock


\bibitem[\protect\citeauthoryear{Haddon}{Haddon}{2015}]%
        {haddon2015children}
\bibfield{author}{\bibinfo{person}{Leslie Haddon}.}
  \bibinfo{year}{2015}\natexlab{}.
\newblock \showarticletitle{Children’s critical evaluation of parental
  mediation}.
\newblock \bibinfo{journal}{\emph{Cyberpsychology: Journal of Psychosocial
  Research on Cyberspace}} \bibinfo{volume}{9}, \bibinfo{number}{1}
  (\bibinfo{year}{2015}), \bibinfo{pages}{2}.
\newblock


\bibitem[\protect\citeauthoryear{Hammond, M{\"u}ller, Carpendale, Bibok, and
  Liebermann-Finestone}{Hammond et~al\mbox{.}}{2012}]%
        {hammond2012effects}
\bibfield{author}{\bibinfo{person}{Stuart~I Hammond}, \bibinfo{person}{Ulrich
  M{\"u}ller}, \bibinfo{person}{Jeremy~IM Carpendale},
  \bibinfo{person}{Maximilian~B Bibok}, {and} \bibinfo{person}{Dana~P
  Liebermann-Finestone}.} \bibinfo{year}{2012}\natexlab{}.
\newblock \showarticletitle{The effects of parental scaffolding on
  preschoolers' executive function.}
\newblock \bibinfo{journal}{\emph{Developmental Psychology}}
  \bibinfo{volume}{48}, \bibinfo{number}{1} (\bibinfo{year}{2012}).
\newblock


\bibitem[\protect\citeauthoryear{Hana and Catherine}{Hana and
  Catherine}{2020}]%
        {eukids2020}
\bibfield{author}{\bibinfo{person}{Machackova Hana} {and}
  \bibinfo{person}{Blaya Catherine}.} \bibinfo{year}{2020}\natexlab{}.
\newblock \bibinfo{title}{Children’s experiences with cyberhate}.
\newblock
\newblock
\urldef\tempurl%
\url{https://www.lse.ac.uk/media-and-communications/assets/documents/research/eu-kids-online/reports}
\showURL{%
\tempurl}


\bibitem[\protect\citeauthoryear{Hashish, Bunt, and Young}{Hashish
  et~al\mbox{.}}{2014}]%
        {hashish2014involving}
\bibfield{author}{\bibinfo{person}{Yasmeen Hashish}, \bibinfo{person}{Andrea
  Bunt}, {and} \bibinfo{person}{James~E Young}.}
  \bibinfo{year}{2014}\natexlab{}.
\newblock \showarticletitle{Involving children in content control: a
  collaborative and education-oriented content filtering approach}. In
  \bibinfo{booktitle}{\emph{Proceedings of the SIGCHI Conf. on Human
  Factors in Computing Systems}}. \bibinfo{pages}{1797--1806}.
\newblock


\bibitem[\protect\citeauthoryear{Kidron and Rudkin}{Kidron and Rudkin}{2017}]%
        {kidron2017digital}
\bibfield{author}{\bibinfo{person}{Beeban Kidron} {and}
  \bibinfo{person}{Angharad Rudkin}.} \bibinfo{year}{2017}\natexlab{}.
\newblock \showarticletitle{Digital Childhood: Addressing childhood development
  milestones in the digital environment.}
\newblock  (\bibinfo{year}{2017}).
\newblock


\bibitem[\protect\citeauthoryear{Ko, Choi, Yang, Lee, and Lee}{Ko
  et~al\mbox{.}}{2015}]%
        {ko2015familync}
\bibfield{author}{\bibinfo{person}{Minsam Ko}, \bibinfo{person}{Seungwoo Choi},
  \bibinfo{person}{Subin Yang}, \bibinfo{person}{Joonwon Lee}, {and}
  \bibinfo{person}{Uichin Lee}.} \bibinfo{year}{2015}\natexlab{}.
\newblock \showarticletitle{FamiLync: facilitating participatory parental
  mediation of adolescents' smartphone use}. In
  \bibinfo{booktitle}{\emph{Proceedings of the 2015 ACM Int'l Joint
  Conf. on Pervasive and Ubiquitous Computing}}.
  \bibinfo{pages}{867--878}.
\newblock


\bibitem[\protect\citeauthoryear{Kopp}{Kopp}{1982}]%
        {kopp1982antecedents}
\bibfield{author}{\bibinfo{person}{Claire~B Kopp}.}
  \bibinfo{year}{1982}\natexlab{}.
\newblock \showarticletitle{Antecedents of self-regulation: a developmental
  perspective.}
\newblock \bibinfo{journal}{\emph{Developmental psychology}}
  \bibinfo{volume}{18}, \bibinfo{number}{2} (\bibinfo{year}{1982}),
  \bibinfo{pages}{199}.
\newblock


\bibitem[\protect\citeauthoryear{Kumar, Naik, Devkar, Chetty, Clegg, and
  Vitak}{Kumar et~al\mbox{.}}{2017}]%
        {kumar2017no}
\bibfield{author}{\bibinfo{person}{Priya Kumar},
  \bibinfo{person}{Shalmali~Milind Naik}, \bibinfo{person}{Utkarsha~Ramesh
  Devkar}, \bibinfo{person}{Marshini Chetty}, \bibinfo{person}{Tamara~L Clegg},
  {and} \bibinfo{person}{Jessica Vitak}.} \bibinfo{year}{2017}\natexlab{}.
\newblock \showarticletitle{'No Telling Passcodes Out Because They're Private'
  Understanding Children's Mental Models of Privacy and Security Online}.
\newblock \bibinfo{journal}{\emph{Proceedings of the ACM on Human-Computer
  Interaction}} \bibinfo{volume}{1}, \bibinfo{number}{CSCW}
  (\bibinfo{year}{2017}), \bibinfo{pages}{1--21}.
\newblock


\bibitem[\protect\citeauthoryear{Kuppens and Ceulemans}{Kuppens and
  Ceulemans}{2019}]%
        {kuppens2019parenting}
\bibfield{author}{\bibinfo{person}{Sofie Kuppens} {and} \bibinfo{person}{Eva
  Ceulemans}.} \bibinfo{year}{2019}\natexlab{}.
\newblock \showarticletitle{Parenting styles: A closer look at a well-known
  concept}.
\newblock \bibinfo{journal}{\emph{Journal of child and family studies}}
  \bibinfo{volume}{28}, \bibinfo{number}{1} (\bibinfo{year}{2019}),
  \bibinfo{pages}{168--181}.
\newblock


\bibitem[\protect\citeauthoryear{Lee}{Lee}{2007}]%
        {lee2007boundary}
\bibfield{author}{\bibinfo{person}{Charlotte~P Lee}.}
  \bibinfo{year}{2007}\natexlab{}.
\newblock \showarticletitle{Boundary negotiating artifacts: Unbinding the
  routine of boundary objects and embracing chaos in collaborative work}.
\newblock \bibinfo{journal}{\emph{Computer Supported Cooperative Work (CSCW)}}
  \bibinfo{volume}{16}, \bibinfo{number}{3} (\bibinfo{year}{2007}),
  \bibinfo{pages}{307--339}.
\newblock


\bibitem[\protect\citeauthoryear{Lee and Chae}{Lee and Chae}{2007}]%
        {lee2007children}
\bibfield{author}{\bibinfo{person}{Sook-Jung Lee} {and}
  \bibinfo{person}{Young-Gil Chae}.} \bibinfo{year}{2007}\natexlab{}.
\newblock \showarticletitle{Children's Internet use in a family context:
  Influence on family relationships and parental mediation}.
\newblock \bibinfo{journal}{\emph{Cyberpsychology \& behavior}}
  \bibinfo{volume}{10}, \bibinfo{number}{5} (\bibinfo{year}{2007}),
  \bibinfo{pages}{640--644}.
\newblock


\bibitem[\protect\citeauthoryear{Lewis and Wharton}{Lewis and Wharton}{1997}]%
        {lewis1997cognitive}
\bibfield{author}{\bibinfo{person}{Clayton Lewis} {and}
  \bibinfo{person}{Cathleen Wharton}.} \bibinfo{year}{1997}\natexlab{}.
\newblock \showarticletitle{Cognitive walkthroughs}.
\newblock In \bibinfo{booktitle}{\emph{Handbook of human-computer
  interaction}}. \bibinfo{publisher}{Elsevier}, \bibinfo{pages}{717--732}.
\newblock


\bibitem[\protect\citeauthoryear{Light, Burgess, and Duguay}{Light
  et~al\mbox{.}}{2018}]%
        {light2018walkthrough}
\bibfield{author}{\bibinfo{person}{Ben Light}, \bibinfo{person}{Jean Burgess},
  {and} \bibinfo{person}{Stefanie Duguay}.} \bibinfo{year}{2018}\natexlab{}.
\newblock \showarticletitle{The walkthrough method: An approach to the study of
  apps}.
\newblock \bibinfo{journal}{\emph{New media \& society}} \bibinfo{volume}{20},
  \bibinfo{number}{3} (\bibinfo{year}{2018}), \bibinfo{pages}{881--900}.
\newblock


\bibitem[\protect\citeauthoryear{Livingstone, Haddon, G{\"o}rzig, and
  {\'O}lafsson}{Livingstone et~al\mbox{.}}{2011}]%
        {livingstone2011risks}
\bibfield{author}{\bibinfo{person}{Sonia Livingstone}, \bibinfo{person}{Leslie
  Haddon}, \bibinfo{person}{Anke G{\"o}rzig}, {and} \bibinfo{person}{Kjartan
  {\'O}lafsson}.} \bibinfo{year}{2011}\natexlab{}.
\newblock \showarticletitle{Risks and safety on the internet: the perspective
  of European children: full findings and policy implications from the EU Kids
  Online survey of 9-16 year olds and their parents in 25 countries}.
\newblock  (\bibinfo{year}{2011}).
\newblock


\bibitem[\protect\citeauthoryear{Livingstone and Helsper}{Livingstone and
  Helsper}{2008}]%
        {livingstone2008parental}
\bibfield{author}{\bibinfo{person}{Sonia Livingstone} {and}
  \bibinfo{person}{Ellen~J Helsper}.} \bibinfo{year}{2008}\natexlab{}.
\newblock \showarticletitle{Parental mediation of children's internet use}.
\newblock \bibinfo{journal}{\emph{Journal of broadcasting \& electronic media}}
  \bibinfo{volume}{52}, \bibinfo{number}{4} (\bibinfo{year}{2008}),
  \bibinfo{pages}{581--599}.
\newblock


\bibitem[\protect\citeauthoryear{Mascheroni, Livingstone, and
  Staksrud}{Mascheroni et~al\mbox{.}}{2015}]%
        {livingstone2015framework}
\bibfield{author}{\bibinfo{person}{Giovanna Mascheroni}, \bibinfo{person}{Sonia
  Livingstone}, {and} \bibinfo{person}{Elisabeth Staksrud}.}
  \bibinfo{year}{2015}\natexlab{}.
\newblock \showarticletitle{Developing a framework for researching children’s
  online risks and opportunities in Europe}.
\newblock  (\bibinfo{date}{12} \bibinfo{year}{2015}).
\newblock


\bibitem[\protect\citeauthoryear{Meeus, Beyens, Geusens, Sodermans, and
  Beullens}{Meeus et~al\mbox{.}}{2018}]%
        {meeus2018managing}
\bibfield{author}{\bibinfo{person}{Anneleen Meeus}, \bibinfo{person}{Ine
  Beyens}, \bibinfo{person}{Femke Geusens}, \bibinfo{person}{An~Katrien
  Sodermans}, {and} \bibinfo{person}{Kathleen Beullens}.}
  \bibinfo{year}{2018}\natexlab{}.
\newblock \showarticletitle{Managing positive and negative media effects among
  adolescents: Parental mediation matters—But not always}.
\newblock \bibinfo{journal}{\emph{Journal of Family Communication}}
  \bibinfo{volume}{18}, \bibinfo{number}{4} (\bibinfo{year}{2018}),
  \bibinfo{pages}{270--285}.
\newblock


\bibitem[\protect\citeauthoryear{Mensah and Kuranchie}{Mensah and
  Kuranchie}{2013}]%
        {mensah2013influence}
\bibfield{author}{\bibinfo{person}{Monica~Konnie Mensah} {and}
  \bibinfo{person}{Alfred Kuranchie}.} \bibinfo{year}{2013}\natexlab{}.
\newblock \showarticletitle{Influence of parenting styles on the social
  development of children}.
\newblock \bibinfo{journal}{\emph{Academic Journal of Interdisciplinary
  Studies}} \bibinfo{volume}{2}, \bibinfo{number}{3} (\bibinfo{year}{2013}),
  \bibinfo{pages}{123--123}.
\newblock


\bibitem[\protect\citeauthoryear{Mesch}{Mesch}{2009}]%
        {mesch2009parental}
\bibfield{author}{\bibinfo{person}{Gustavo~S Mesch}.}
  \bibinfo{year}{2009}\natexlab{}.
\newblock \showarticletitle{Parental mediation, online activities, and
  cyberbullying}.
\newblock \bibinfo{journal}{\emph{CyberPsychology \& Behavior}}
  \bibinfo{volume}{12}, \bibinfo{number}{4} (\bibinfo{year}{2009}),
  \bibinfo{pages}{387--393}.
\newblock


\bibitem[\protect\citeauthoryear{Miller}{Miller}{1995}]%
        {miller1995wordnet}
\bibfield{author}{\bibinfo{person}{George~A Miller}.}
  \bibinfo{year}{1995}\natexlab{}.
\newblock \showarticletitle{WordNet: a lexical database for English}.
\newblock \bibinfo{journal}{\emph{Commun. ACM}} \bibinfo{volume}{38},
  \bibinfo{number}{11} (\bibinfo{year}{1995}), \bibinfo{pages}{39--41}.
\newblock


\bibitem[\protect\citeauthoryear{Moilanen, Rasmussen, and
  Padilla-Walker}{Moilanen et~al\mbox{.}}{2015}]%
        {moilanen2015bidirectional}
\bibfield{author}{\bibinfo{person}{Kristin~L Moilanen},
  \bibinfo{person}{Katie~E Rasmussen}, {and} \bibinfo{person}{Laura~M
  Padilla-Walker}.} \bibinfo{year}{2015}\natexlab{}.
\newblock \showarticletitle{Bidirectional associations between self-regulation
  and parenting styles in early adolescence}.
\newblock \bibinfo{journal}{\emph{Journal of research on adolescence}}
  \bibinfo{volume}{25}, \bibinfo{number}{2} (\bibinfo{year}{2015}),
  \bibinfo{pages}{246--262}.
\newblock


\bibitem[\protect\citeauthoryear{Mollborn and Fomby}{Mollborn and
  Fomby}{2020}]%
        {mollborn2020making}
\bibfield{author}{\bibinfo{person}{Stefanie Mollborn} {and}
  \bibinfo{person}{Paula Fomby}.} \bibinfo{year}{2020}\natexlab{}.
\newblock \showarticletitle{Making Sense of Kids' Technology Use}.
\newblock  (\bibinfo{year}{2020}).
\newblock


\bibitem[\protect\citeauthoryear{Nathanson}{Nathanson}{1999}]%
        {nathanson1999identifying}
\bibfield{author}{\bibinfo{person}{Amy~I Nathanson}.}
  \bibinfo{year}{1999}\natexlab{}.
\newblock \showarticletitle{Identifying and explaining the relationship between
  parental mediation and children's aggression}.
\newblock \bibinfo{journal}{\emph{Communication Research}}
  \bibinfo{volume}{26}, \bibinfo{number}{2} (\bibinfo{year}{1999}),
  \bibinfo{pages}{124--143}.
\newblock


\bibitem[\protect\citeauthoryear{Nathanson}{Nathanson}{2013}]%
        {nathanson2013media}
\bibfield{author}{\bibinfo{person}{Amy~I Nathanson}.}
  \bibinfo{year}{2013}\natexlab{}.
\newblock \showarticletitle{Media and the family context}.
\newblock In \bibinfo{booktitle}{\emph{The Routledge Int'l handbook of
  children, adolescents and media}}. \bibinfo{publisher}{Routledge},
  \bibinfo{pages}{325--332}.
\newblock


\bibitem[\protect\citeauthoryear{Nikken and Jansz}{Nikken and Jansz}{2014}]%
        {nikken2014developing}
\bibfield{author}{\bibinfo{person}{Peter Nikken} {and} \bibinfo{person}{Jeroen
  Jansz}.} \bibinfo{year}{2014}\natexlab{}.
\newblock \showarticletitle{Developing scales to measure parental mediation of
  young children's internet use}.
\newblock \bibinfo{journal}{\emph{Learning, Media and technology}}
  \bibinfo{volume}{39}, \bibinfo{number}{2} (\bibinfo{year}{2014}),
  \bibinfo{pages}{250--266}.
\newblock


\bibitem[\protect\citeauthoryear{Ofcom}{Ofcom}{2019}]%
        {st2019}
\bibfield{author}{\bibinfo{person}{Ofcom}.} \bibinfo{year}{2019}\natexlab{}.
\newblock \bibinfo{title}{Why children spend time online}.
\newblock
\newblock
\urldef\tempurl%
\url{https://www.ofcom.org.uk/about-ofcom/latest/media/media-releases/2019/why-children-spend-time-online}
\showURL{%
\tempurl}


\bibitem[\protect\citeauthoryear{Ofcom.org}{Ofcom.org}{2019}]%
        {ofcom2019}
\bibfield{author}{\bibinfo{person}{Ofcom.org}.}
  \bibinfo{year}{2019}\natexlab{a}.
\newblock \bibinfo{title}{children and parents: media use and attitudes report
  2019}.
\newblock
\newblock
\urldef\tempurl%
\url{https://www.ofcom.org.uk/__data/assets/pdf_file/0023/190616/children-media-use-attitudes-2019-report.pdf}
\showURL{%
\tempurl}


\bibitem[\protect\citeauthoryear{Ofcom.org}{Ofcom.org}{2020}]%
        {ofcom2020}
\bibfield{author}{\bibinfo{person}{Ofcom.org}.}
  \bibinfo{year}{2020}\natexlab{b}.
\newblock \bibinfo{title}{Ofcom Children’s Media Lives: Life in Lockdown}.
\newblock
\newblock
\urldef\tempurl%
\url{https://www.ofcom.org.uk/__data/assets/pdf_file/0024/200976/cml-life-in-lockdown-report.pdf}
\showURL{%
\tempurl}


\bibitem[\protect\citeauthoryear{ofcom.org}{ofcom.org}{2016}]%
        {ofcom2016}
\bibfield{author}{\bibinfo{person}{ofcom.org}.}
  \bibinfo{year}{2016}\natexlab{}.
\newblock \bibinfo{title}{Children and parents: media use and attitude report}.
\newblock
\newblock
\urldef\tempurl%
\url{https://www.ofcom.org.uk/research-and-data/media-literacy-research/childrens/children-parents-nov16}
\showURL{%
\tempurl}


\bibitem[\protect\citeauthoryear{ofcom.org}{ofcom.org}{2017}]%
        {ofcom2017}
\bibfield{author}{\bibinfo{person}{ofcom.org}.}
  \bibinfo{year}{2017}\natexlab{}.
\newblock \bibinfo{title}{Children and parents: media use and attitude report}.
\newblock
\newblock
\urldef\tempurl%
\url{https://www.ofcom.org.uk/__data/assets/pdf_file/0020/108182/children-parents-media-use-attitudes-2017.pdf}
\showURL{%
\tempurl}


\bibitem[\protect\citeauthoryear{Paulsen, Platt, Huettel, and Brannon}{Paulsen
  et~al\mbox{.}}{2012}]%
        {paulsen2012risk}
\bibfield{author}{\bibinfo{person}{David~J Paulsen}, \bibinfo{person}{Michael~L
  Platt}, \bibinfo{person}{Scott~A Huettel}, {and} \bibinfo{person}{Elizabeth~M
  Brannon}.} \bibinfo{year}{2012}\natexlab{}.
\newblock \showarticletitle{From risk-seeking to risk-averse: the development
  of economic risk preference from childhood to adulthood}.
\newblock \bibinfo{journal}{\emph{Frontiers in psychology}}
  \bibinfo{volume}{3} (\bibinfo{year}{2012}), \bibinfo{pages}{313}.
\newblock


\bibitem[\protect\citeauthoryear{Pine and Nash}{Pine and Nash}{2002}]%
        {pine2002dear}
\bibfield{author}{\bibinfo{person}{Karen~J Pine} {and} \bibinfo{person}{Avril
  Nash}.} \bibinfo{year}{2002}\natexlab{}.
\newblock \showarticletitle{Dear Santa: The effects of television advertising
  on young children}.
\newblock \bibinfo{journal}{\emph{Int'l Journal of Behavioral
  Development}} \bibinfo{volume}{26}, \bibinfo{number}{6}
  (\bibinfo{year}{2002}), \bibinfo{pages}{529--539}.
\newblock


\bibitem[\protect\citeauthoryear{Pinter, Wisniewski, Xu, Rosson, and
  Caroll}{Pinter et~al\mbox{.}}{2017}]%
        {pinter2017adolescent}
\bibfield{author}{\bibinfo{person}{Anthony~T Pinter}, \bibinfo{person}{Pamela~J
  Wisniewski}, \bibinfo{person}{Heng Xu}, \bibinfo{person}{Mary~Beth Rosson},
  {and} \bibinfo{person}{Jack~M Caroll}.} \bibinfo{year}{2017}\natexlab{}.
\newblock \showarticletitle{Adolescent online safety: Moving beyond formative
  evaluations to designing solutions for the future}. In
  \bibinfo{booktitle}{\emph{Proceedings of the 2017 Conf. on Interaction
  Design and Children}}. \bibinfo{pages}{352--357}.
\newblock


\bibitem[\protect\citeauthoryear{Research}{Research}{2020}]%
        {PEW2020}
\bibfield{author}{\bibinfo{person}{Pew Research}.}
  \bibinfo{year}{2020}\natexlab{}.
\newblock \bibinfo{title}{Parenting Children in the Age of Screens}.
\newblock
\newblock
\urldef\tempurl%
\url{https://www.pewresearch.org/internet/2020/07/28/parenting-children-in-the-age-of-screens/}
\showURL{%
\tempurl}


\bibitem[\protect\citeauthoryear{Reynolds and Darden}{Reynolds and
  Darden}{1971}]%
        {reynolds1971mutually}
\bibfield{author}{\bibinfo{person}{Fred~D Reynolds} {and}
  \bibinfo{person}{William~R Darden}.} \bibinfo{year}{1971}\natexlab{}.
\newblock \showarticletitle{Mutually adaptive effects of interpersonal
  communication}.
\newblock \bibinfo{journal}{\emph{Journal of Marketing Research}}
  \bibinfo{volume}{8}, \bibinfo{number}{4} (\bibinfo{year}{1971}),
  \bibinfo{pages}{449--454}.
\newblock


\bibitem[\protect\citeauthoryear{Ryan and Deci}{Ryan and Deci}{2000}]%
        {ryan2000self}
\bibfield{author}{\bibinfo{person}{Richard~M Ryan} {and}
  \bibinfo{person}{Edward~L Deci}.} \bibinfo{year}{2000}\natexlab{}.
\newblock \showarticletitle{Self-determination theory and the facilitation of
  intrinsic motivation, social development, and well-being.}
\newblock \bibinfo{journal}{\emph{American psychologist}} \bibinfo{volume}{55},
  \bibinfo{number}{1} (\bibinfo{year}{2000}), \bibinfo{pages}{68}.
\newblock


\bibitem[\protect\citeauthoryear{Shin and Lwin}{Shin and Lwin}{2017}]%
        {shin2017does}
\bibfield{author}{\bibinfo{person}{Wonsun Shin} {and} \bibinfo{person}{May~O
  Lwin}.} \bibinfo{year}{2017}\natexlab{}.
\newblock \showarticletitle{How does “talking about the Internet with
  others” affect teenagers’ experience of online risks? The role of active
  mediation by parents, peers, and school teachers}.
\newblock \bibinfo{journal}{\emph{New Media \& Society}} \bibinfo{volume}{19},
  \bibinfo{number}{7} (\bibinfo{year}{2017}).
\newblock


\bibitem[\protect\citeauthoryear{Smahel, MacHackova, Mascheroni, Dedkova,
  Staksrud, Olafsson, Livingstone, and Hasebrink}{Smahel et~al\mbox{.}}{2020}]%
        {smahel2020eu}
\bibfield{author}{\bibinfo{person}{David Smahel}, \bibinfo{person}{Hana
  MacHackova}, \bibinfo{person}{Giovanna Mascheroni}, \bibinfo{person}{Lenka
  Dedkova}, \bibinfo{person}{Elisabeth Staksrud}, \bibinfo{person}{Kjartan
  Olafsson}, \bibinfo{person}{Sonia Livingstone}, {and} \bibinfo{person}{Uwe
  Hasebrink}.} \bibinfo{year}{2020}\natexlab{}.
\newblock \showarticletitle{EU Kids Online 2020: Survey results from 19
  countries}.
\newblock  (\bibinfo{year}{2020}).
\newblock


\bibitem[\protect\citeauthoryear{Sonia and Julia}{Sonia and Julia}{2017}]%
        {harm2017}
\bibfield{author}{\bibinfo{person}{Livingstone Sonia} {and}
  \bibinfo{person}{Davidson Julia}.} \bibinfo{year}{2017}\natexlab{}.
\newblock \bibinfo{title}{Children’s online activities, risks and safety}.
\newblock
\newblock


\bibitem[\protect\citeauthoryear{Strub}{Strub}{1989}]%
        {strub1989theory}
\bibfield{author}{\bibinfo{person}{Harry Strub}.}
  \bibinfo{year}{1989}\natexlab{}.
\newblock \showarticletitle{The theory of panoptical control: Bentham's
  panopticon and Orwell's Nineteen Eighty-Four}.
\newblock \bibinfo{journal}{\emph{Journal of the History of the Behavioral
  Sciences}} \bibinfo{volume}{25}, \bibinfo{number}{1} (\bibinfo{year}{1989}),
  \bibinfo{pages}{40--59}.
\newblock


\bibitem[\protect\citeauthoryear{Symons, Ponnet, Emmery, Walrave, and
  Heirman}{Symons et~al\mbox{.}}{2017}]%
        {symons2017parental}
\bibfield{author}{\bibinfo{person}{Katrien Symons}, \bibinfo{person}{Koen
  Ponnet}, \bibinfo{person}{Kathleen Emmery}, \bibinfo{person}{Michel Walrave},
  {and} \bibinfo{person}{Wannes Heirman}.} \bibinfo{year}{2017}\natexlab{}.
\newblock \showarticletitle{Parental knowledge of adolescents’ online content
  and contact risks}.
\newblock \bibinfo{journal}{\emph{Journal of youth and adolescence}}
  \bibinfo{volume}{46}, \bibinfo{number}{2} (\bibinfo{year}{2017}),
  \bibinfo{pages}{401--416}.
\newblock


\bibitem[\protect\citeauthoryear{Temple}{Temple}{1997}]%
        {temple1997cognitive}
\bibfield{author}{\bibinfo{person}{Christine~M Temple}.}
  \bibinfo{year}{1997}\natexlab{}.
\newblock \showarticletitle{Cognitive neuropsychology and its application to
  children}.
\newblock \bibinfo{journal}{\emph{Journal of Child Psychology and Psychiatry}}
  \bibinfo{volume}{38}, \bibinfo{number}{1} (\bibinfo{year}{1997}),
  \bibinfo{pages}{27--52}.
\newblock


\bibitem[\protect\citeauthoryear{Unicef}{Unicef}{2020}]%
        {st2020}
\bibfield{author}{\bibinfo{person}{Unicef}.} \bibinfo{year}{2020}\natexlab{}.
\newblock \bibinfo{title}{Rethinking screen-time in the time of COVID-19}.
\newblock
\newblock
\urldef\tempurl%
\url{https://www.unicef.org/globalinsight/stories/rethinking-screen-time-time-covid-19}
\showURL{%
\tempurl}


\bibitem[\protect\citeauthoryear{Valkenburg, Krcmar, Peeters, and
  Marseille}{Valkenburg et~al\mbox{.}}{1999}]%
        {valkenburg1999developing}
\bibfield{author}{\bibinfo{person}{Patti~M Valkenburg}, \bibinfo{person}{Marina
  Krcmar}, \bibinfo{person}{Allerd~L Peeters}, {and} \bibinfo{person}{Nies~M
  Marseille}.} \bibinfo{year}{1999}\natexlab{}.
\newblock \showarticletitle{Developing a scale to assess three styles of
  television mediation:“Instructive mediation,”“restrictive mediation,”
  and “social coviewing”}.
\newblock \bibinfo{journal}{\emph{Journal of broadcasting \& electronic media}}
  \bibinfo{volume}{43}, \bibinfo{number}{1} (\bibinfo{year}{1999}),
  \bibinfo{pages}{52--66}.
\newblock


\bibitem[\protect\citeauthoryear{Valkenburg, Piotrowski, Hermanns, and
  De~Leeuw}{Valkenburg et~al\mbox{.}}{2013}]%
        {valkenburg2013developing}
\bibfield{author}{\bibinfo{person}{Patti~M Valkenburg},
  \bibinfo{person}{Jessica~Taylor Piotrowski}, \bibinfo{person}{Jo Hermanns},
  {and} \bibinfo{person}{Rebecca De~Leeuw}.} \bibinfo{year}{2013}\natexlab{}.
\newblock \showarticletitle{Developing and validating the perceived parental
  media mediation scale: A self-determination perspective}.
\newblock \bibinfo{journal}{\emph{Human Communication Research}}
  \bibinfo{volume}{39}, \bibinfo{number}{4} (\bibinfo{year}{2013}),
  \bibinfo{pages}{445--469}.
\newblock


\bibitem[\protect\citeauthoryear{Wang, Zhao, and Shadbolt}{Wang
  et~al\mbox{.}}{2019}]%
        {wang2019concerns}
\bibfield{author}{\bibinfo{person}{Ge Wang}, \bibinfo{person}{Jun Zhao}, {and}
  \bibinfo{person}{Nigel Shadbolt}.} \bibinfo{year}{2019}\natexlab{}.
\newblock \showarticletitle{What concerns do Chinese parents have about their
  children's digital adoption and how to better support them?}
\newblock \bibinfo{journal}{\emph{arXiv preprint arXiv:1906.11123}}
  (\bibinfo{year}{2019}).
\newblock


\bibitem[\protect\citeauthoryear{Wildemuth}{Wildemuth}{2016}]%
        {wildemuth2016applications}
\bibfield{author}{\bibinfo{person}{Barbara~M Wildemuth}.}
  \bibinfo{year}{2016}\natexlab{}.
\newblock \bibinfo{booktitle}{\emph{Applications of social research methods to
  questions in information and library science}}.
\newblock \bibinfo{publisher}{ABC-CLIO}.
\newblock


\bibitem[\protect\citeauthoryear{Wisniewski, Ghosh, Xu, Rosson, and
  Carroll}{Wisniewski et~al\mbox{.}}{2017}]%
        {wisniewski2017parental}
\bibfield{author}{\bibinfo{person}{Pamela Wisniewski},
  \bibinfo{person}{Arup~Kumar Ghosh}, \bibinfo{person}{Heng Xu},
  \bibinfo{person}{Mary~Beth Rosson}, {and} \bibinfo{person}{John~M Carroll}.}
  \bibinfo{year}{2017}\natexlab{}.
\newblock \showarticletitle{Parental Control vs. Teen Self-Regulation: Is there
  a middle ground for mobile online safety?}. In
  \bibinfo{booktitle}{\emph{Proceedings of the 2017 ACM Conf. on Computer
  Supported Cooperative Work and Social Computing}}. \bibinfo{pages}{51--69}.
\newblock


\bibitem[\protect\citeauthoryear{Wisniewski, Jia, Xu, Rosson, and
  Carroll}{Wisniewski et~al\mbox{.}}{2015}]%
        {wisniewski2015preventative}
\bibfield{author}{\bibinfo{person}{Pamela Wisniewski}, \bibinfo{person}{Haiyan
  Jia}, \bibinfo{person}{Heng Xu}, \bibinfo{person}{Mary~Beth Rosson}, {and}
  \bibinfo{person}{John~M Carroll}.} \bibinfo{year}{2015}\natexlab{}.
\newblock \showarticletitle{" Preventative" vs." Reactive" How Parental
  Mediation Influences Teens' Social Media Privacy Behaviors}. In
  \bibinfo{booktitle}{\emph{Proceedings of the 18th ACM Conf. on Computer
  Supported Cooperative Work \& Social Computing}}. \bibinfo{pages}{302--316}.
\newblock


\bibitem[\protect\citeauthoryear{Zaman and Nouwen}{Zaman and Nouwen}{2016}]%
        {zaman2016parental}
\bibfield{author}{\bibinfo{person}{Bieke Zaman} {and} \bibinfo{person}{Marije
  Nouwen}.} \bibinfo{year}{2016}\natexlab{}.
\newblock \showarticletitle{Parental controls: advice for parents, researchers
  and industry}.
\newblock \bibinfo{journal}{\emph{EU Kids Online}}.
\newblock


\bibitem[\protect\citeauthoryear{Zhao, Wang, Dally, Slovak, Edbrooke-Childs,
  Van~Kleek, and Shadbolt}{Zhao et~al\mbox{.}}{2019}]%
        {zhao2019make}
\bibfield{author}{\bibinfo{person}{Jun Zhao}, \bibinfo{person}{Ge Wang},
  \bibinfo{person}{Carys Dally}, \bibinfo{person}{Petr Slovak},
  \bibinfo{person}{Julian Edbrooke-Childs}, \bibinfo{person}{Max Van~Kleek},
  {and} \bibinfo{person}{Nigel Shadbolt}.} \bibinfo{year}{2019}\natexlab{}.
\newblock \showarticletitle{I make up a silly name' Understanding Children's
  Perception of Privacy Risks Online}. In \bibinfo{booktitle}{\emph{Proceedings
  of the 2019 CHI Conf. on Human Factors in Computing Systems}}.
  \bibinfo{pages}{1--13}.
\newblock


\end{thebibliography}
\end{document}